\pdfoutput=1
\RequirePackage{ifpdf}
\ifpdf 
\documentclass[pdftex]{sigma}
\else
\documentclass{sigma}
\fi

\newtheorem{Theorem}{Theorem}[section]
\newtheorem{Problem}[Theorem]{Problem}
\numberwithin{equation}{section}
\DeclareMathOperator{\supp}{supp}

\begin{document}

\newcommand{\arXivNumber}{1407.2124}

\allowdisplaybreaks

\renewcommand{\PaperNumber}{085}

\FirstPageHeading

\ShortArticleName{The Ongoing Impact of Modular Localization on Particle Theory}

\ArticleName{The Ongoing Impact of Modular Localization\\
on Particle Theory}

\Author{Bert SCHROER~$^{\dag\ddag}$}

\AuthorNameForHeading{B.~Schroer}

\Address{$^\dag$~CBPF, Rua Dr. Xavier Sigaud 150, 22290-180 Rio de Janeiro, Brazil}
\EmailD{\href{mailto:schroer@cbpf.br}{schroer@cbpf.br}}

\Address{$^\ddag$~Institut f\"{u}r Theoretische Physik, FU-Berlin, Arnimallee 14, 14195 Berlin, Germany}

\ArticleDates{Received July 05, 2013, in f\/inal form July 28, 2014; Published online August 13, 2014}

\Abstract{Modular localization is the concise conceptual formulation of causal localization in the setting of local
quantum physics. \looseness=-1
Unlike QM it does not refer to individual operators but rather to ensembles of observables which share the same
localization region, as a~result it explains the probabilistic aspects of QFT in terms of the impure KMS nature arising
from the local restriction of the pure vacuum.
Whereas it played no important role in the perturbation theory of low spin particles, it becomes indispensible for
interactions which involve higher spin $s\geq1$ f\/ields, where is leads to the replacement of the operator (BRST) gauge
theory setting in Krein space by a~new formulation in terms of stringlocal f\/ields in Hilbert space.
The main purpose of this paper is to present new results which lead to a~rethinking of important issues of the Standard
Model concerning massive gauge theories and the Higgs mechanism.
We place these new f\/indings into the broader context of ongoing conceptual changes within QFT which already led to new
nonperturbative constructions of models of integrable QFTs.
It is also pointed out that modular localization does not support ideas coming from string theory, as extra dimensions
and Kaluza--Klein dimensional reductions outside quasiclassical approximations.
Apart from hologarphic projections on null-surfaces, holograhic relations between QFT in dif\/ferent spacetime dimensions
violate the causal completeness property, this includes in particular the Maldacena conjecture.
Last not least, modular localization sheds light onto unsolved problems from QFT's distant past since it reveals that
the Einstein--Jordan conundrum is really an early harbinger of the Unruh ef\/fect.}


\Classification{47L15; 81P05; 81P40; 81R40; 81T05; 81T40}

\rightline{\it To the memory of Hans-J\"{u}rgen Borchers (1926--2011)}

\vspace{-2mm}

\section{Introduction}

The course of quantum f\/ield theory (QFT) was to a~large extend determined by four important conceptual conquests: its
1926 discovery by Pascual Jordan in the aftermath of what in recent times has been referred to as the
\textit{Einstein--Jordan conundrum}~\cite{Du-Ja,E-J} (a fascinating dispute between Einstein and Jordan), the discovery
of renormalized perturbation in the context of quantum electrodynamics (QED) after world war II, the nonperturbative
insights into the particle-f\/ield relation initiated in the Lehmann--Symanzik--Zimmermann (LSZ) work on scat\-te\-ring theory
which subsequently was derived from f\/irst principles~\cite{Haag} and the extension of gauge theory leading up to the
Standard Model and to the present research in particle physics.

Especially the nonperturbative derivation of time-dependent scattering theory from the foundational causal locality
properties of QFT in conjunction with the dif\/f\/icult task to describe strong interactions led to the f\/irst solid insights
into particle theory outside the range of perturbation theory.
One of those nonperturbative results was the rigorous derivation of the particle analog of the Kramers--Kronig dispersion
relations.
This in turn led to their subsequent successful experimental test which was very important for the continued trust in
QFT's foundational causality principle at the new high energy scales.
These results in turn encouraged a~third project: the particle-based on-shell formulations known as the $S$-matrix
bootstrap and Mandelstam's more analytic formulation in terms of auxiliary two-variable representations of elastic
scattering amplitudes~\cite{Man}.
These on-shell projects as well as their dual model and string theory (ST) successors were less successful, to put it
mildly.
The later gauge theory of the Standard Model resulted from an extension of the QED quantization ideas.
Despite its undeniable success it was not able to prevent the ascend of ST, partially because ST withdrew from problems
of high energy particle theory with the (never fulf\/illed) promise to solve the foundational problem of quantum gravity
at the length scale of the Planck distance.

One of the most remarkable innovative contributions of the 60s was Gell-Mann's idea of quark conf\/inement and his later
attempts to pose it as a~conceptual challenge for QCD.
Although its interpretative addition to QCD turned out to be remarkably consistent, its derivation as a~ma\-the\-matical
consequence of that theory resisted all attempts undertaken during the 50 years of its existence.
The reason it is mentioned in this introduction goes beyond historical completeness; the new stringlocal (SLF) Hilbert
space setting of Yang--Mills couplings sheds new light on this old problem (Section~\ref{Section3}).

Despite conceptual weaknesses, the Standard Model has remained the phenomenologically most inclusive and successful
particle description.
Its theoretical foundations date back to the early 70s and the experimental progress during more than 4~decades did
not require any signif\/icant theoretical changes.
In particular its central theoretical idea that masses of vector mesons and of particles with which they interact are
generated by a~spontaneous symmetry breaking (the Higgs mechanism), which led to last year's physics Nobel prize,
remained in its original form in which it appeared f\/irst in the papers of Higgs and Englert.
This is surprising since during its 40 years history several authors have cast valid doubts about its consistency with
the principles of QFT.

The main point of the present work consists in the proposal of a~new idea which extends \textit{renormalized
perturbation theory in a~Hilbert space} setting to f\/ields of higher spin $s\geq1$.
At this point it is important to
remind the reader that the gauge theoretic formulation of interactions of vector mesons with matter f\/ields (massless and
massive abelian and nonabelian interactions) uses an indef\/inite metric Krein space and unphysical ghost operators.
The loss of a~Hilbert space description is the price one has to pay for maintaining the formalism of renormalized
perturbation theory \textit{in terms of pointlike fields} for interactions involving higher spin f\/ields with
$s\geq1$. The new setting maintains the Hilbert space description but leaves it up to the causal localization principles to
determine the tightest localization which is still consistent with the Hilbert space setting of quantum theory.
The answer is that one never has to go beyond stringlocal f\/ields.
This clash between localization and the Hilbert space structure and pointlike localization of f\/ields is a~quantum
phenomenon which has no counterpart in classical theory; it explains why Lagrangian quantization of $s\geq1$ inevitably
leads to Krein space formulations.

\looseness=-1
The reformulation of gauge theory in terms of interactions between stringlocal f\/ields (SLF) in Hilbert space is much
more than window-dressing: it extends the range of gauge theories beyond the construction of local observables to the
inclusion of (necessarily stringlocal) physical matter f\/ields and opens a~realistic chance to understand conf\/inement as
a~physical property of a~model and not just an auxiliary metaphoric idea for exploring its physical consequences.
The SLF inverts the relation between massless gauge theories and their massive counterparts; instead of
considering models involving massless vector mesons as simpler than their massive counterparts, the SLF setting describes
the massless models (QED, QCD) as massless limits of QFTs with a~complete particle interpretation (validity of LSZ
scattering theory) since the problem of scattering of stringlocal charged matter in QED remained on a~level of a~recipe
(rather than of a~spacetime explanation); not to mention this issue of gluon and quark conf\/inement.

It is not surprising that such a~paradigm shift also leads to a~change of the ``Higgs mechanism of spontaneous symmetry
breaking'' which in the new setting is simply the renormalizable coupling of a~massive vector meson to real (Hermitian)
scalar matter and the \textit{postulated} Mexican hat potential (which served as the formal description of the symmetry
breaking Higgs mechanism) is now the \textit{result} of interaction terms which the implementation of the SLF locality
principle \textit{induces} from the iteration of the f\/irst-order interaction.
In particular there is no generation of masses of vector mesons by a~Higgs mechanism.
Our f\/indings show that interactions of massive vector mesons with matter can be consistently described within the Hilbert
space setting of QT without referring to a~mass-generating Higgs mechanism.
Though f\/ields of massive vector mesons are always accompanied by scalar f\/ields, their inexorable presence (``intrinsic
escorts'') does not lead to additional degrees of freedom.
Their presence is the result of by the positivity of Hilbert space which for interactions of massive $s\geq1$ turns out
to have very strong consequences; it does not only lead to stringlocal instead of pointlocal f\/ields, but it also
generates~$s$ additional escort f\/ields of lower spin.

\looseness=-1
The scalar escort for $s=1$ has many properties of a Higgs
f\/ield except that it does not add degrees of freedom and therefore can only
explain the LHC experimental result in terms of a~bound state. On the other
hand a Higgs coupling in the new setting is simply described by a~coupling
of a massive vector meson to a~Hermitian f\/ield~$H$ (``chargeless QED'');
but the principles of QFT certainly exclude the idea that massive
vector mesons owe their mass to spontaneous symmetry breaking of gauge
symmetry in scalar QED.

It is interesting that this is not the f\/irst time the Higgs mechanism came under critical scrutiny.
In fact in the work of the Z\"{u}rich group from the beginnings of the 90s~\cite{Aste, Scharf} based on the
operator BRST formulation of (the simpler case) of massive vector mesons-matter interactions it was shown that the
Mexican hat potential is not the def\/ining interaction but rather the second order outcome from the implementation of the
BRST gauge invariance on a~f\/irst-order interaction which results from a~transcription into the Krein space setting of
a~nonrenormali\-zable f\/irst-order $A^{\rm P}\cdot A^{\rm P}H$ coupling, where $A_{\mu}^{\rm P}$ is the massive Proca potential of the
vector meson.
In all cases to which the new SLF Hilbert space formulation was applied, these earlier results from BRST gauge theory
were conf\/irmed, although the details of the SLF Hilbert space setting are dif\/ferent and the range of this method is larger.
Results similar to those in Section~\ref{Section3} are contained in~\cite{Schroer} and furthergoing results
about nonabelian couplings will be contained in a~forthcoming joint work with J.~Mund~\cite{M-S}.

The ongoing paradigmatic change also suggests to recall other critical ideas which were around at the time of the Higgs
paper but whose content was lost in the maelstrom of time.
On such idea is the Schwinger--Swieca charge screening which was suggested by Schwinger~\cite{schwinger} and proven by
Swieca~\cite{Sw}.
It states that abelian massive vector meson couplings possess (in addition to the conserved current of complex f\/ields
which leads to the global counting charge) also an identically conserved Maxwell current (the divergence of the f\/ield
strength) whose charge vanishes (``is screened'').
For charge neutral matter f\/ields, as in the Higgs model, this is the only current.

\looseness=1
It would be possible to present these results directly without embedding them into their natural conceptual
surroundings from which they emerged.
But since these conceptual developments are only known to a~very small circle of theoreticians, and also since the new
emerging picture about what QFT can and should still achieve is as important as its ongoing impact on gauge theory and
the Higgs mechanism, the special results on SLF will be placed into a~larger context.
To this enlarged setting also implies a~foundational critique of ST, in particular because without a~clear delimitations
between the incorrect use of the word ``string'' in ST and its foundational deployment in SLF this could lead to misconceptions.
In addition there is no better constructive
use of an incorrect but widespread known theory than to use it for showing in
what way a subtle concept as quantum causal localization has been
misunderstood.

\looseness=-1
The starting point is what is nowadays referred to as Local Quantum Physics (LQP)~\cite{Haag}.
This is a~way of looking at QFT in which quantum f\/ields are considered as generators of loca\-li\-zed operator algebras;
they ``coordinatize'' local nets of algebras in analogy to coordinates in geo\-met\-ry which coordinatize a~given model geometry.
This is quite dif\/ferent from the way one looks at classical f\/ield theories where, e.g.,
Maxwell's electromagnetic f\/ield strength has an intrinsic mea\-ning.
Such an ``individuality'' of f\/ields gets lost in QFT beyond quasiclassical approximations.
Experimentalists do not observe hadronic f\/ields; what is being
measured are hadronic particles entering or leaving a~collision area.
But unlike quantum mechanics (QM) particles have no direct relation to individual f\/ields, rather a~particle carrying
a~certain super\-selected charge is related to a~whole f\/ield class which consists of f\/ields carrying the same charge and
belong to the same localization class (relative locality).
The justif\/ication for this point of view results from the fact that these f\/ields ``interpolate'' the same particle.
For more details about the subtle f\/ield-particle relations see~\cite{Haag}.

\looseness=-1
The f\/irst contact between the Tomita--Takesaki modular theory of operator algebras and quantum physics came from
quantum statistical mechanics, to be more precise from the formulation of statistical mechanics directly in the
thermodynamic inf\/inite volume limit (``open systems'')~\cite{Haag}.
The important observation was that the prerequisites for the application of the T-T theory (an algebra $\mathcal{A}$ and
a~state vector $\Omega$ on which it acts cyclic and separating, see later) is always fulf\/illed in statistical mechanics.
As a~consequence the two ``modular operators'' $\Delta^{i\tau}$ and~$J$ have a~physical interpretation in statistical
mechanics where~$\Delta$ is the so-called KMS ope\-ra\-tor (the thermodynamic limit of the Gibbs operator) and $J$ is
a~anti-unitary ref\/lection which maps the algebra $\mathcal{A}$ into its commutant (the thermal ``shadow world'').
The essential step which opened the use of the T-T theory in LQP was the realization of the validity of the
Reeh--Schlieder theorem for the pair ($\mathcal{A}(\mathcal{O}),\Omega$) where $\mathcal{A}(\mathcal{O})$~is an algebra localized in the
spacetime region $\mathcal{O}$ and~$\Omega$ is the vacuum state.
The Reeh--Schlieder theorem is closely related to a~very singular form of entanglement of the vacuum with respect to
a~subdivision of the global algebra~$\mathcal{A}$ into~$\mathcal{A}(\mathcal{O})$ of the region~$\mathcal{O}$ and that of its
causal complement~$\mathcal{O}^{\prime}$.
This singular entanglement is related to the fact that although the algebra and its causal complement commute with each other,
the global Hilbert space does not tensor-factorize.
In contrast to the entanglement of quantum mechanical particle states which can be measured in terms of
quantum-optical methods, the ef\/fects of the impurity of the $\mathcal{A}(\mathcal{O})$-restricted vacuum (Unruh ef\/fect, Hawking
radiation) entanglement are numerically so tiny that they may remain always below what can be measured.
Nevertheless the vacuum polarization through localization is behind almost most physical manifestations of QFT, from
analytic on-shell behavior (as the particle crossing property of the $S$-matrix and formfactors) to the Unruh
ef\/fect~\cite{Sewell, Unruh} and the area law for localization entropy~\cite{causal}.

\looseness=-1
A historically particularly interesting manifestation of the statistical mechanics nature of the state resulting from
the local restriction of the vacuum is the so-called Einstein--Jordan conundrum which similar to the Unruh ef\/fect shows
that the subvolume f\/luctuations of a~reduced vacuum state in the simplest QFT (the chiral abelian current model) are
indistinguishable from those of in a~thermal statistical mechanic state of the kind which Einstein used for his purely
theo\-retical argument for the corpuscular nature of photons.
If these facts would have been correctly identif\/ied, the history of the probability concept in quantum theory may have
taken another turn.
The algebra of local observables $\mathcal{A}(\mathcal{O})$ is an ensemble of observables to which the restriction of the pure
vacuum state generates an impure KMS state.
It is reasonable to use the name \textit{physical states} only for f\/inite energy states and to reserve the terminology
\textit{observable} to operators which are localized in some compact spacetime region and obey Einstein causality.
Since the statistical mechanics-like KMS property holds not only for the vacuum but also for the \textit{restriction of
all physical states} to local observables, the probabilistic aspect resulting from the from the ensemble of observables
localized in a~spacetime region $\mathcal{O}$ is a~generic intrinsic property of all physical states in QFT (which
Einstein would have accepted).
In contrast, for individual observables in QM one needs to invoke Born's probability interpretation\footnote{Here we use
this terminology in the textbook sense of Born's localization probability density $\left\vert \psi(x)\right\vert^{2}$
which results from declaring a~particular operator $\vec{x}_{\rm op}$ to be a~position operator.} which refers to
a~``Gedanken''-ensemble related to repeated measurements (to which Einstein had his philosophical objections).
The best chance to obtain a~deeper understanding of the QFT/QM relation is in the context of integrable models where
actual particle creation (through collisions) is absent but vacuum polarization as the inexorable epiphenomenon of
modular localization remains.

The SLF setting is an outgrowth of the solution of the problem of the QFT behind Wigner's 1939 third positive energy
representation class (the massless inf\/inite spin representations).
In that case all f\/ields associated to the representation are stringlocal, not just potentials of ge\-ne\-ral f\/ield
strengths.
The resulting matter is ``noncompact'' in an intrinsic sense~\cite{MSY}.
It has all the properties ascribed by astrophysicists to dark matter, i.e.~it is inert and its arena of manifestations
are galaxies and not earthly high energy laboratories~\cite{dark}.

The paper is organized as follows.
The next section presents a~foundational critique of ST in which already the terminology reveals the misconception
of the meaning of quantum causal localization; part of this misunderstanding results from confusing Born's localization of
wave functions based on the spectral decomposition of the (non-intrinsic) position operator and part is due to a~``picture
puzzle'' resulting from the fact that the 10 component supersymmetric chiral current algebra is a~representation of
a~corresponding irreducible $C^{\ast}$-algebra of oscillators on which there also exists a~positive energy
representation of the 10-dimensional highly reducible so-called superstring representation of the Poincar\'e group.

\looseness=-1
Having sharpened one's view on causal localization, the presentation then moves to \textit{modular localization} which
is the most appropriate conceptual as well as mathematical formulation of quantum causal localization.
Its application to Wigner's positive energy representation theory of the Poincar\'{e} group led to the QFT of the
inf\/inite spin representation which is generated by irreducibly string-localized covariant f\/ields.
Irreducibly stringlocal interacting f\/ields result from the interaction of reducibly stringlocal free f\/ields.
Section~\ref{Section3} and its subsections are the heart piece of a~new SLF approach to
perturbative QFT which includes higher spin interactions.
Its relation to the existing BRST gauge setting is explained, and its already mentioned critical view of the Higgs
mechanism is presented in detail.
The SLF setting sheds new light on the conf\/inement problem and reduces it to a~computational problem involving
perturbative resummations.

In the last section known results about existence proofs of integrable models are used to formulate conjectures about
how modular theory may help to obtain a~mathematical control of existence problems of QFT.
The section also explains how the particle crossing property arises from modular wedge-localization.

Our f\/indings support the title and the content of an important contribution by the late Hans-J\"{u}rgen Borchers in the
millennium edition of Journal of Mathematical Physics~\cite{Bo} which reads: ``Revolutionizing quantum f\/ield theory with
Tomita--Takesaki's modular theory''.
With all reservations about misuses of the word ``revolution'' in particle physics, this paper is a~comprehensive account
of the role of modular operator theory in LQP, and its title is a~premonition of the present progress which is driven~by
concepts coming from modular localization.
LQP owns Borchers many of the concepts coming from modular operator theory; for this reason it is very appropriate to
dedicate the present article to his memory.

\vspace{-1mm}

\section{Anomalous conformal dimensions, particle spectra\\ and crossing properties}\label{Section2}

A large part of the conceptual derailment of string theory can be understood without invoking the subtleties of modular
localization.
This will be the subject of the following two subsections.

The principle of \textit{modular localization} becomes however essential for the correct foundational understanding of
the particle crossing property. This is important for a~new formulation of a~constructive on-shell project based on the
correct crossing property which replaces Mandelstam's attempt and is compatible with the principles of Haag's local
quantum physics.
This will be taken up in Section~\ref{Section4}.

\subsection{Quantum mechanical- versus causal-localization}

Since part of the misunderstandings in connection with ST have to some extend their origin in confusing ``Born
localization'' in QM with the causal localization in QFT, it may be helpful to review their signif\/icant
dif\/ferences~\cite{interface}.

It is well-known since Wigner's 1939 description of relativistic particles~\cite{Haag} in terms of irreducible positive
energy representations of the Poincar\'{e} group that \textit{there are no $4$-component covariant operators}
$x_{\rm op}^{\mu}$; in fact the impossibility to describe relativistic particles in terms of quantizing a~classical
relativistic particle action (or to achieve this in any other quantum mechanical setting) was one of the reasons which led to Wigner's
representation theoretical construction of relativistic wave function spaces.
The rather simple argument against covariant selfadjoint $x_{\rm op}^{\mu}$ follows from the non-existence translationally
covariant spectral projectors~$E$ which are consistent with the positive energy condition and fulf\/ill with spacelike
orthogonality
\begin{gather*}
  \vec{x}_{\rm op}=\int\vec{x}dE_{\vec{x}},
  \qquad
  R\subset\mathbb{R}^{3}\rightarrow E(R),
\\
  U(a)E(R)U(a)^{-1}=E(R+a),
  \qquad
  E(R)E(R^{\prime})=0
  \quad
  \text{for}
  \quad
  R\times R^{\prime},
\\
  \left(E(R)\psi,U(a)E(R)\psi\right) =\left(\psi,E(R)E(R+a)U(a)\psi \right) =0,
\end{gather*}
where the second line expresses translational covariance and orthogonality of projections for spacelike separated
regions.
In the third line we assumed that the translation shifts $E(R)$ spacelike with respect to itself.
But since $U(a)\psi$ is analytic in $\mathbb{R}^{4}+iV^{+}$ ($V^{+}$ forward light cone) as a~result of the spectrum
condition, $ \Vert E(R)\psi \Vert^{2}=0$ for all~$R$ and $\psi$ which implies $E(R)\equiv0$, i.e.~covariant
position operators do not exist.

The ``Born probability'' of QM results from Born's proposal to interpret the absolute square $ \vert
\psi(\vec{x},t) \vert^{2}$ of the spectral decomposition $\psi(\vec{x},t)$ of state vectors with respect to the
spectral resolution of the position operator $\vec{x}_{\rm op}(t)$
as the probability density to f\/ind the particle at time~$t$ at the position $\vec{x}$.
Its use as a~localization probability density to f\/ind an individual particle in a~pure state at a~prescribed position
became the beginning of one of longest lasting philosophical disputes in QM which Einstein entered through his famous
saying: ``God does not play dice''.

In QFT in Haag's LQP formulation this problem does not exist since, as previously mentioned, its objects of interests
are not global position operators in individual quantum mechanical systems, but rather ensembles of causally localized
operators which share the same spacetime localization, i.e.,~which belong to the spacetime-indexed algebras
$\mathcal{A}(\mathcal{O}$) of Haag's LQP (next section).
As pointed out before this leads to a~completely intrinsic notion of probability.

Traditionally the causal localization of QFT enters the theory with the (graded) spacelike commutation (Einstein
causality) of pointlike localized covariant f\/ields in Minkowski spacetime.
There are very good reasons to pass to another slightly more general, but in a~subtle sense also more specif\/ic
formulation of QFT, namely to Haag's \textit{local quantum physics} (LQP) in which the f\/ields play a~more auxiliary role
of (necessary singular) generators of local algebras\footnote{To be more precise they are operator-valued Schwartz
distributions whose smearing with $\mathcal{O}$-supported test functions are (generally unbounded) operators af\/f\/iliated
with a~weakly closed operator algebra $\mathcal{A}(\mathcal{O})$.}.
In analogy to coordinates in geometry as there are inf\/initely many such generators which generate the same local net of
algebras as dif\/ferent coordinates which describe the same geometry.
As in Minkowski spacetime geometry these ``f\/ield coordinates'' can be chosen globally, i.e.~for the generation of the full
net of local algebras.

In this more conceptual LQP setting it is easier to express the \textit{full} content of causal loca\-li\-za\-tion in
a~precise operational setting.
It includes not only the Einstein causality for spacelike separated local observables, but also a~timelike aspect of
causal localization, namely the equality of an $\mathcal{O}$-localized operator algebra $\mathcal{A}(\mathcal{O})$ with that of
its causal completion $\mathcal{O}^{\prime\prime}$
\begin{gather*}
\mathcal{A}(\mathcal{O})=\mathcal{A}(\mathcal{O}^{\prime\prime}), \qquad \text{causal completeness},
\\
\mathcal{A}(\mathcal{O}^{\prime})= \mathcal{A}(\mathcal{O})^{\prime}, \qquad \text{Haag~duality}.
\end{gather*}
Here \looseness=-1 $\mathcal{O}^{\prime}$ denotes the causal complement (consisting of all points which
are spacelike with respect to $\mathcal{O}$) and $\mathcal{O}^{\prime\prime}= (\mathcal{O}^{\prime})^\prime$ is the causal completion. Haag duality is stronger than Einstein causality (which results
from replacing~$=$ by~$\subset$). The notation $\mathcal{A}^\prime$ for the commutant of $\mathcal{A}$ is
standard in the theory of operator algebras. The causal completeness
requirement corresponds to the classical causal propagation property.
The advantage of the LQP formulation over the use of pointlike f\/ields should
be obvious.
A~more specif\/ic picture of a~failure of causal completeness due to a~mismatch of degrees of freedom results if one compares the def\/inition of a
local algebra localized in a convex spacetime region $\mathcal{O}$ obtained in two
dif\/ferent ways, on the one hand as an
intersection of wedge algebras (outer approximation def\/ining the causal completion) and on the other hand as a~union of arbitrary small
double cones (inner approximation).
In case the region is not causally complete the inner approximation is smaller that
the outer one $\mathcal{A}(\mathcal{O}) := \mathcal{A}_{\rm in}(\mathcal{O}) \varsubsetneqq \mathcal{A}_{\rm out}(\mathcal{O}) =: \mathcal{A}(\mathcal{O}'')$.
In this case there is a~serious physical problem since there are degrees of
freedom which have entered the causal completion from ``nowhere'' (``poltergeist'' degrees of freedom).

Whereas both causality requirements are \textit{formal} attributes of Lagrangian quantization (hyperbolic propagation),
they have to be added in ``axiomatic'' settings based on \textit{mathematically controlled} (and hence neither Lagrangian
nor functional) formulations~\cite{H-S}.
Their violations for subalgebras $\mathcal{A}(\mathcal{O})$ as a~result of too many phase space degrees\footnote{For the notion of
phase space degree of freedoms see~\cite{ABL,Bu-Wi, H-Sw}.} of freedom leads to physically undesirable
ef\/fects, which among other things prevents the mathematical AdS-CFT correspondence (last subsection) to admit a~physical
interpretation on both sides of the correspondence (i.e.~one side is always unphysical).

Violations of Haag duality for disconnected or multiply connected regions have interesting physical
consequences in connection with either the superselection sectors associated with observable algebras, or with the QFT
Aharonov--Bohm ef\/fect for doubly connected spacetime algebras for the free quantum Maxwell f\/ield with possible
generalizations to multiply connected spacetime regions in higher spin ($m=0$, $s\geq1$)
representations~\cite{nonlocal, charge}.

The LQP formulation of QFT is naturally related to the Tomita--Takesaki modular theory of operator algebras; its general
validity for spacetime localized algebras in QFT is a~direct result of the Reeh--Schlieder property~\cite{Haag} for
localized algebras $\mathcal{A}(\mathcal{O})$, $\mathcal{O}^{\prime\prime} \subset\mathbb{R}^{4}$ (next section).

It is important to understand that \textit{quantum mechanical localization is not cogently related with spacetime}.
A~linear chain of oscillators simply does not care about the dimension of space in which it is pictured; in fact it does
not even care if it is related to spacetime at all, or whether it refers to some internal space to which spacetime
causality concepts are not applicable.
The modular localization on the other hand is \textit{imprinted} on causally local quantum matter, it is a~totally
\textit{holistic} property of such matter.
As life cannot be explained in terms of the chemical composition of a~living body, localization does not follow from the
mathematical description of the global oscillators (annihilation/creation operators) in a~global algebra.
These oscillators are the same in QM and QFT; free f\/ield oscillator variables $a(p)$, $a^{\ast}(p)$ which obey the
oscillator commutation relations do not know whether they will be used in order to def\/ine Schr\"{o}dinger f\/ields or free
covariant local quantum f\/ields.

It is the holistic \textit{modular localization principle} which imprints the causal properties of Min\-kow\-ski spacetime
(including the spacetime dimension) on operator algebras and thus determines in which way the irreducible system of
oscillators will be used in the process of localization~\cite{Ho-Wa}; in QFT there is no abstract quantum matter as
there is in QM; rather localization becomes an inseparable part of it.
Contrary to a~popular belief, this holistic aspect of QFT (in contrast to classical theory and Born's localization in
QM) \textit{does not permit an embedding of a~lower-dimensional theory into a~higher-dimensional} one, neither is its
inversion (Kaluza--Klein reduction, branes) possible.
To be more specif\/ic, the price for compressing a~QFT onto a~timelike hypersurface~\cite{Borchers} is the loss of
physical content namely one looses the important timelike causal completeness property due to an abundance of degrees of
freedom.
One may study such restrictions as laboratories for testing problems of mathematical physics, but they have no relevance
for particle physics.
This does however not include projections onto null-surfaces which \textit{reduce the cardinality of degrees of freedom}
(unlike the K-K reductions and AdS-CFT holographic isomorphisms\footnote{A concise mathematical description of this
phenomenon (but without a~presentation of the physical consequences) can be found in~\cite{Reh}.} which maintain it).
We will return to this issue in later parts of the paper.
There has been an attempt by Mack~\cite{Mack2, Mack1} to encode the overpopulation of degrees of freedom into
a~generalization of internal symmetries, but this does not seem to make the situation acceptable.
If one only uses such situations as a~mathematical trick (e.g.\
for doing calculations of an ${\rm AdS}_{5}$ QFT on the side of the overpopulated ${\rm CFT}_{4}$ theory before returning again) and
not in the sense of Maldacena (allegedly relating two physical theories) this generates no harm.

One problem in reading articles or books on ST is that it is sometimes dif\/f\/icult to decide which localization they have
in mind.
When e.g.\
Polchinski~\cite{Polch} uses the relativistic particle action~$\sqrt{ds^{2}}$ as a~trailer for the introduction of the
Nambu-Goto minimal surface action $\sqrt{A}~($with~$A$ being the quadratic surface analog of the line element
$ds^{2})$ for a~description of ST, it is not clear why he does this.
These Lagrangians lead to relativistic classical equation of motion
but the classical particle Lagrangian is known to possess no associated
relativistic quantum theory.

The Polyakov action~$A$ can be formally written in terms of the potential of an $n$-component chiral current
\begin{gather*}
\int d\sigma d\tau\sum\limits_{\xi=\sigma,\tau}\partial_{\xi}X_{\mu}(\sigma,\tau)g^{\mu\nu}\partial^{\xi}X_{\mu}(\sigma,\tau),
\qquad
X=\text{potential of conformal current} \  j.
\end{gather*}
However the quantum theory related to the Nambu--Goto action has nothing to do with its square (see later).
The widespread use of the letter~$X$ for the potential of the multicomponent chiral current is very treacherous
since it suggests that Polchinski's incorrect quantum mechanical reading of the classical $\sqrt{ds^2}$ has led to the incorrect idea that the action of
a~multi-component $d=1+1$ massless f\/ield describes in some way a~covariant string embedded into a~higher-dimensional
Minkowski spacetime (a~kind of relativistic analog of a~linear chain of oscillators into a~higher-dimensional~QM).

If the quantized~$X$ of the Polyakov action would really describe a~covariant spacetime string, one could forget about
the N-G square root action and take the Polyakov action for the construction of an embedded string.
But this cannot work since the principle of modular localization simply contradicts the idea that a~lower-dimensional
QFT can be embedded into a~higher-dimensional one.
In particular an $n$-component chiral conformal QFT cannot be \textit{embedded} as a~``source'' theory into a~QFT which is
associated with a~representation of the Poincar\'{e} group acting on the $n$-component inner symmetry space (the ``target''
space) of a~conformal f\/ield theo\-ry.
The $C^{\ast}$-algebra generated by the oscillators contained in a~$d=10$ supersymmetric chiral current model carries
a~representation of the $d=1+1$ Moebius group and possesses a~(unitarily inequivalent) representation which carries
a~positive energy representation of the Poincar\'{e} group; but from this one cannot infer the existence of a~spacetime
``embedding'' of a~1-dimensional chiral theory localized on the compactif\/ied lightray into a~10-dimensional QFT.

String theorists gave a~correct proof of this group theoretic fact~\cite{Brower}, but in order to construct an $S$-matrix
it takes more than group theory.
\textit{In fact the global oscillator algebra admits at least two inequivalent representations}: one on which the
M\"{o}bius group acts and in which it is possible to construct pointlike M\"{o}bius covariant f\/ields, and the other on
which the mentioned unique 10-dimensional highly reducible representation of the Poincar\'{e} group acts.
The easiest way to see that the representations are dif\/ferent is to notice that the multi-component charge spectrum is
continuous whereas the corresponding Poincar\'{e} momentum spectrum has mass gaps.
In addition the embedding picture would incorrectly suggest that the object is a~spacetime string and not an inf\/inite
component pointlike wave function or quantum f\/ield as required by a~f\/inite spin/helicity positive energy
representation\footnote{Only the zero mass inf\/inite spin representation leads to string-localization~\cite{MSY}.}.
The group theoretic theorem cannot be used in an on-shell $S$-matrix approach. To construct an $S$-matrix one needs more
than just group representation theory of the Poincar\'e group.
Admittedly the mentioned group theoretic theorem is somewhat surprising since \textit{it is the only known irreducible
algebra which leads to a~discrete mass/spin tower} (no admixture of a~continuous energy-momentum spectrum coming from
multiparticle states).

\looseness=-1
Often a~better conceptual understanding is obtained by generalizing a~special situation.
Instead of an irreducible algebra associated with a~chiral current theory one may ask whether an \textit{internal}
symmetry space of a~f\/inite component quantum f\/ield can (i.e.~not indices referring to spinor/tensor components of
f\/ields) carry the representation of a~noncompact group.
In classical theories this is always possible, whereas in QFT one would certainly not expect this in $d>1+1$ models.
For theories with mass gaps this is the result of a~deep theorem about the possible superselection structure of
observable LQP algebras~\cite{Haag}; there are good reasons to believe that this continues to hold for the charge
structure in theories containing massless f\/ields~\cite{Bu}.
A~necessary prerequisite is the existence of continuously many superselected charges as in the case of abelian current models.
By def\/inition this is the class of non-rational chiral models.
Apart from the multicomponent abelian current model almost nothing is known about this class; so the problem of whether
the ``target spaces'' of such models can accommodate unitary representations of noncompact groups (i.e.~the question
whether the above theorem about unitary representations on multicomponent current algebras is a~special case of a~more
general phenomenon) remains open.

\looseness=-1
A rather trivial illustration of a~classical theory on whose index space a~Poincar\'{e} group acts without the existence of
a~quantum counterpart is the afore-mentioned relativistic classical mechanics.
As covariant classical theories may not possess a~quantum counterpart, there are also strong indications about the
existence of QFTs which cannot be pictured as the quantized version of covariant classical f\/ields\footnote{This goes
also in the opposite direction: there are many known $d=1+1$ integrable models which have no Lagrangian description.}.
The best way of presenting the group theoretical theorem of the string theorists is to view it in a~historical context
as the (presently only known) solution of the 1932 Majorana project~\cite{Maj}.
Majorana was led to his idea about the possible natural existence of inf\/inite component relativistic f\/ields by the
${\rm O}(4,2)$ group theoretical description of the nonrelativistic hydrogen spectrum.
We take the liberty to formulate it here in a~more modern terminology.

\begin{Problem}[Majorana~\cite{Maj}] Find an irreducible algebraic structure which carries a~infinite-component positive energy one-particle
representation of the Poincar\'{e} group $($an ``infinite component wave equation''$)$.
\end{Problem}

\looseness=-1
Majorana's own search, as well as that for the so-called ``dynamic inf\/inite component f\/ield equation'' by a~group in the
60s (Fronsdal, Barut, Kleinert, \dots; see appendix of~\cite{To}) consisted in looking for irreducible group algebras of
noncompact extensions of the Lorentz group (``dyna\-mi\-cal groups'').
No acceptable solution was ever found within such a~setting.
The only known solution is the above superstring representation which results from an irreducible oscillator algebra of
the $n=10$ supersymmetric Polyakov model.
The positive energy property of its particle content (and the absence of components of Wigner's ``inf\/inite spin''
components) secures the pointlike localizability of this ``superstring representation'' (too late to change this
unfortunate terminology).

\looseness=-1
Sometimes the confusions about localization did not directly enter the calculations of string theorists but rather remained in the
interpretation.
A~poignant illustration is the calculation of the (graded) commutator of string f\/ields in~\cite{Lowe,Martinec}.
Apart from the technical problem that inf\/inite component f\/ields can not be tempered distribution (since the piling up of
free f\/ields over one point with ever increasing masses and spins leads to a~diverging short distance scaling behavior
which requires to project onto f\/inite mass subspaces), the graded commutator is pointlike.
This was precisely the result of that calculation; but the authors presented their result as ``the (center) point on
a~string''.
Certainly this uncommon distributional behavior has no
relation with the idea of spacetime strings; at most one may speak about a~quantum mechanical chain of oscillators in
``inner space'' (over a~localization point).
The memory of the origin of ST from an \textit{irreducible} oscillator algebra is imprinted in the fact that the degrees
of freedoms used for the representation of the Poincar\'{e} group do not exhaust the oscillator degrees of freedom,
there remain degrees of freedom which interconnect the representations in the $(m,s)$ tower, i.e.~which prevent that the
oscillator algebra representation is only a~direct sum of wave function spaces.
But the localization properties reside fully in these wave function spaces and, as a~result of the absence of Wigner's
inf\/inite spin representations, the localization is pointlike.
This is precisely what the above-mentioned authors found, but why did they not state this clearly.

ST led to the bizarre suggestion that we are living in an (dimensionally reduced) target space of an (almost)
unique\footnote{Up to a~f\/inite number of M-theoretic modif\/ications.} 10-dimensional chiral conformal theory.
A~related but at f\/irst sight more appearing idea is the dimensional reduction which was proposed in the early
days of quantum theory by Kaluza and Klein.
Both authors illustrated their idea in classical/semiclassical f\/ield theory\footnote{Also ``branes'' were only explained
in the context of quasiclassical approximations.}; nobody ever established its validity in a~full-f\/ledged QFT (e.g.\
on the level of its correlation functions) was never established.
There is a~good reason for this since the idea is in conf\/lict with the foundational causal localization property. Unlike
Born localization in QM, modular localization is an intrinsic property; the concept of matter in LQP cannot be separated
from spacetime, it is rather coupled to its dimensionality through the spacetime dependent notion of ``degrees of
freedom''. As explained in the previous section this is closely related to causality (the ``causal completenes property'')
where it was pointed out that e.g.\
the mathematical AdS-CFT algebraic isomorphism converts a~physical QFT on one side of the correspondence into a physically unacceptable model on the ``wrong'' side.
This does however not exclude the possibility that it may be easier to do computations on the other side of the
isomorphism and transform
the computed result back to the physical side.

The idea of the use of variable spacetime dimensions in QFT (the ``epsilon expansion'') goes back to Ken Wilson who used
it as a~method (a technical trick) for computing anomalous dimensions (critical indices) of scalar f\/ields.
But whereas the method gave reasonable results for critical indices
of scalar f\/ields, this is certainly not the case for $s>1$ matter; as already Wigner's classif\/ication of particles and their related free f\/ields show, the appearance of
changing ``little groups'' prevents an analytic dependence.

Our criticism of the dual model and ST is two-fold, on the one hand the reader will be reminded that the meromorphic
crossing properties of the dual model, although not related to particle theory, represent a~rigorous property of conformal correlations after passing to their Mellin transform.
The poles in these variables occur at the scale dimensions of composites which appear in global operator expansions of
two conformal covariant f\/ields.
In this formal game of producing crossing symmetric functions through Mellin transforms the spacetime dimensionality
does not play any role; any conformal QFT leads to such a~dual model function and that found by Veneziano belongs to a~chiral current
model.
A~special distinguished spectrum appears if one performs a~Mellin-transforms on the correlations of the 10-component current model
whose oscillator algebra carries the unitary positive energy ``superstring representation''  of the Poincar\'{e} group
(the previously mentioned only known solution of the Majorana problem).
In this case the $(m,s)$ Poincar\'{e} spectrum is proportional to the dimensional spectrum $(d,s)$ of composites which appear
in the global operator expansion of the anomalous dimension-carrying sigma f\/ields which are associated to the chiral current model.

The second criticism of the dual model/ST is that scattering amplitudes cannot be
meromorphic in the Mandelstam variables; in integrable models they are meromorphic in the rapidities.
The best way to understand the physical content of particle crossing is to derive it from the analytic formulation of
the KMS property for modular wedge localization.
This does not only reveal the dif\/ference to dual model crossing, but also suggests a~new on-shell construction methods
based on the $S$-matrix which may be capable to replace Mandelstam's approach (Section~\ref{Section3}).

\subsection{The picture puzzle of chiral models and particle spectra}

There are two ways to see the correct mathematical-conceptual meaning of the dual model and~ST.

One uses the ``Mack machine''~\cite{Mack2, Mack1} for the construction of dual models (including the dual model
which Veneziano constructed ``by hand'').
It starts from a~4-point function of \textit{any conformal QFT in any spacetime dimension}.
To maintain simplicity we take the vacuum expectation of four not necessarily equal scalar f\/ields
\begin{gather*}
\left\langle A_{1}(x_{1})A_{2}(x_{2})A_{3}(x_{3})A_{4}(x_{4})\right\rangle.
\end{gather*}
It is one of the specialities of interacting conformal theories that f\/ields have no associated particles with a~discrete
mass, instead they carry (generally a~non-canonical, anomalous, discrete) scale dimensions which are connected with the
nontrivial \textit{center of the conformal covering group}~\cite{integrable}.
It is well known from the pre BPZ~\cite{BPZ} conformal research in the 70s~\cite{Ma-Lu,Sc-Sw} that conformal
theories have converging operator expansions of the type
\begin{gather}
A_{3}(x_{3})A_{4}(x_{4})\Omega=\sum\limits_{k}\int d^{4}z\Delta_{A_{3},A_{4.},C_{k}}(x_{1},x_{2},y)C_{k}(z)\Omega,
\nonumber
\\
\left\langle A_{1}(x_{1})A_{2}(x_{2})A_{3}(x_{3})A_{4}(x_{4})\right\rangle
\
\rightarrow
\
3~\text{dif\/ferent expansions}.
\label{3}
\end{gather}
In \looseness=-1 distinction to the Wilson--Zimmermann short distance expansions, which only converge in an asymptotic sense, these
expansions converge in the sense of state-vector valued Schwartz distributions.
The form of the global 3-point-like expansion coef\/f\/icients is completely f\/ixed in terms of the anomalous scale dimension
spectrum of the participating conformal f\/ields.

It is clear that there are exactly three ways of applying global operator expansions to pairs of operators inside
a~4-point-function~\eqref{3}; they are analogous to the three possible particle pairings in the elastic $S$-matrix which
correspond to the $s$, $t$ and~$u$ in Mandelstam's formulation of crossing.
But beware, this dual model crossing arising from the Mellin transform of conformal correlation has nothing to do with
$S$-matrix particle crossing of Mandelstam's on-shell project.
If duality would have arisen in this context probably nobody would have connected them with the particle crossing in
$S$-matrices and on-shell formfactors.
Veneziano found these relations \cite{Fub, Ven}
by using mathematical properties of Euler beta function Euler beta function; his
construction did not reveal its conformal origin.
Since particle crossing and its conceptual origin in the principles of QFT remained ill-understood (for a~recent
account of its origin from modular localization see~\mbox{\cite{integrable, causal}}), the incorrect identif\/ication of
crossing with Veneziano's duality met little resistance.

The Mellin transform of the 4-point-function is a~meromorphic function in $s$, $t$, $u$ which has f\/irst-order poles at the
numerical values of the anomalous dimensions of those conformal composites which appear in the three dif\/ferent
decompositions of products of conformal f\/ields; they are related by analytic continuation~\cite{Mack2, Mack1}.
To enforce an interpretation of particle masses, one may rescale these dimensionless numbers by the same dimensionfull
number.
However this formal step of calling the scale dimensions of composites particle masses does not change the physical
reality.
Structural analogies in particle physics are worthless without an independent support concerning their physical origin.

The Mack machine to produce dual models (crossing symmetric analytic functions of 3 va\-riab\-les) has no def\/inite relation
to spacetime dimensions; one may start from a~\textit{conformal theory in any spacetime dimension} and end with
a~meromorphic crossing function in Mellin variables.
Calling them Mandelstam variables does not change the conceptual-mathematical reality deletion; one is dealing with two quantum objects whose position in Hilbert
space can hardly be more dif\/ferent than that of scattering amplitudes and conformal correlations.

However, and here we come to the picture-puzzle aspect of ST, one can ask the more mo\-dest question whether one can view
the \textit{dimensional spectrum of composites} in global operator expansions (after multiplication with a~common
dimensionfull $[m^{2}]$ parameter) \textit{as arising from a~positive energy representation of the
Poincar\'{e} group}.
The only such possibility which was found is the previously mentioned 10 component superymmetric chiral current theory
which leads to the well-known superstring representation of the Poincar\'{e} group and constitutes the only known
solution of the Majorana project.
In this way the analogy of the anomalous composite dimensions of the poles in the dual model from the Mack machine to
a~$(m,s)$ mass spectrum is extended to a~genuine particle representation of the Poincar\'{e} group.
But even this lucky circumstance which leads to the superstring representation remains on the level of group theory and
by its very construction cannot be viewed as containing dynamic informations about a~scattering amplitude.

There exists a~presentation which exposes this ``picture-puzzle'' aspect between conformal chiral current models and
Wigner's particle representation properties in an even stronger way: the so-called sigma-model representation.
Schematically it can be described in terms of the following manipulation on abelian chiral currents ($x=$ lightray
coordinate)
\begin{gather}
\partial\Phi_{k}(x)   =j_{k}(x), \qquad \Phi_{k}(x)=\int_{-\infty}^{x} j_{k}(x), \qquad \left\langle
j_{k}(x)j_{l}(x^{\prime})\right\rangle \sim\delta_{k,l}\left(x-x^{\prime}-i\varepsilon\right)^{-2},
\label{ana}
\\
Q_{k}   =\Phi_{k}(\infty),
\qquad
\Psi(x,\vec{q})= :e^{i\vec{q}\vec{\Phi} (x)} :, \qquad \text{carries~$\vec{q}$-charge},
\nonumber
\\
Q_{k}   \simeq P_{k}, \qquad \dim(e^{i\vec{q}\vec{\Phi}(x)})\sim\vec{q}\cdot\vec {q}\simeq p_{\mu}p^{\mu}, \qquad (d_{\rm sd},s)\sim(m,s).
\nonumber
\end{gather}
The f\/irst line def\/ines the \textit{potentials of the current}; it is formally infrared-divergent.
The vacuum sector is instead created by applying the polynomial algebra generated by the infrared convergent current.
In contrast \textit{the exponential sigma field~$\Psi$ is the formal expression for a~covariant superselected
charge-carrying field}. Its symbolic exponential way of writing leads to the \textit{correct correlation functions in
total charge zero correlations} where the correlation functions agree with those computed from
Wick-reordering of
products of sigma model f\/ields $\Psi$, all other correlations of the sigma-model f\/ield vanish (the quotation mark is
meant to indicate this limitation of the Wick ordering).

\looseness=1
The interesting line is the third in~\eqref{ana}, since it expresses a~``mock relation'' with particle physics; the
multi-component continuous charge spectrum of the conformal currents resemble a~continuous momentum spectrum of
a~representation of the Poincar\'{e} group, whereas the spectrum of anomalous scale dimensions (being quadratic in the
charges) is reminiscent the quadratic relation between momenta and particle masses.
The above analogy amounts to a~genuine po\-si\-tive energy representation of the Poincar\'{e} group only in the special case
of a~supersymmetric 10-component chiral current model; it is the before-mentioned solution of the Majorana project.
Its appearance in the Mellin transform of a~conformal correlation bears no relation with an $S$-matrix.
As also mentioned, the shared irreducible abstract oscillator algebra leads to dif\/ferent representations in its
conformal use from that for a~positive energy representation of the Poincar\'{e} group\footnote{The 26 component model
does not appear here because we are interested in localizable representation; only positive energy representations are
localizable.}.
The dif\/ference between the representation leading to the conformal chiral theory and that of the Poincar\'{e} group on
the target space (the superstring representation) prevents the (structurally anyhow impossible) interpretation in terms
of an embedding of QFTs; although there remains a~certain proximity as a~result of the shared oscillator al\-gebra.

The multicomponent $Q_{\mu}$ charge spectrum covers the \textit{full} $\mathbb{R}^{10}$ whereas the $P_{\mu}$ spectrum
of the superstring representation is concentrated on \textit{positive mass hyperboloids}.
The Hilbert space representation of the oscillator algebra from
the Fourier decomposition of the compactif\/ied conformal current model
on which one obtains a realization of the M\"{o}bius
group is not the same as that which leads to the superstring representation
of the Poincar\'{e} group.
Hence presenting the result as an embedding of the chiral ``source theory'' into the 10 component ``target theory'' is
a~misunderstanding caused by the ``picture-puzzle'' aspect
of the sigma f\/ield formulation of the chiral current model.
The representation theoretical dif\/ferences express the dif\/ferent holistic character of the two dif\/ferent localizations
(the target localization being a~direct consequence of the intrinsic localization of positive energy representations of
the Poincar\'{e} group).
This picture-puzzle situation leads to two mathematical questions which will not be further pursued: why does the
positive energy representation of the Poincar\'{e} group only occur when the chiral realization has a~vanishing Virasoro
algebra parameter? And are there other non-rational (continuous set of superselection sectors) chiral models which solve
the Majorana project?

\looseness=1
It should be added that it would be totally misleading to reduce the mathematical/conceptual use of chiral abelian
current models to their role in the solution of the Majorana project of constructing inf\/inite component wave equations.
The chiral $n$-component current models played an important conceptual role in mathematical physics; the so-called
\textit{maximal extensions} of these observable algebras can be classif\/ied by integer lattices, and the possible
superselection sectors of these so extended algebras are classif\/ied in terms of their dual
lattices~\cite{BMT,Do-Xu,Ka-Lo,Sta}.
Interestingly the \textit{selfdual lattices} and their known relation with exceptional f\/inal groups correspond precisely
to the \textit{absence of non-vacuum superselection sectors} (no nontrivial superselected charges) which in turn is
equivalent to the validity of \textit{full} Haag duality (Haag duality also for all multiply-connected
algebras~\cite{nonlocal, charge}).
They constitute the most explicitly constructed nontrivial chiral models.
They shed light on the interplay of discrete group theory and Haag duality (and also on its violation for localization
on disconnected intervals).

\section{Wigner representations and their covariantization}\label{Section3}

Historically the use of the new setting of modular localization started with a~challenge since the days of
Wigner's particle classif\/ication: \textit{find the causal localization of the third Wigner class} (the massless
inf\/inite spin class) of positive energy representations of the Poincar\'{e} group.
Whereas the massive class as well as the zero-mass f\/inite helicity class are pointlike generated, it is not possible to
f\/ind covariant pointlike generating wave functions for this third Wigner class.
The f\/irst representation theoretical argument showing the impossibility of a~pointlike generation dates back
to~\cite{Yold}.
Decades later new ideas about the use of modular localization in connection with integrable models emerged~\cite{AOP}.
This was followed by the concept of modular localization of wave functions in the setting of Wigner's positive energy
representation of the Poincar\'{e} group~\cite{BGL} which led to the introduction of spacelike string-generated f\/ields
in~\cite{MSY}.
These are covariant f\/ields $\Psi(x,e)$, $e$ spacelike unit vector, which are localized on $x+\mathbb{R}_{+}e$ in the sense
that the (graded) commutator vanishes if the full semiinf\/inite strings (and not only their starting points $x)$ are
spacelike separated~\cite{MSY}
\begin{gather}
\left[\Psi(x,e),\Phi(x^{\prime},e^{\prime})\right]_{\text{grad}}=0, \qquad
x+\mathbb{R}_{+}e\rangle\langle~x^{\prime}+\mathbb{R}_{+}e^{\prime}.
\label{string}
\end{gather}
Unlike decomposable stringlike f\/ields (pointlike f\/ields integrated along spacelike half\/lines) such \textit{elementary
stringlike fields} lead to serious problems with respect to the activation of (compactly localized) particle counters.
The decomposable free strings of higher spin potentials (see next subsection) are in an appropriate sense ``milder''.
As pointlike localized f\/ields, free string-localized f\/ields have Fourier transforms which are on-shell (mass-shell).

In the old days~\cite{Weinbook}, inf\/inite spin representations were rejected on the ground that nature does not make use
of them.
But whether in times of dark matter one would uphold such dismissals is questionable, in particular since it turn out
that they have the desired inert/invisibility properties~\cite{dark} which one attributes to dark matter.

Dif\/ferent from pointlike f\/ields, string-localized quantum f\/ields f\/luctuate both in~$x$ as well as in~$e$;\footnote{These
long distance (infrared) f\/luctuations are short distance f\/luctuation in the sense of the asymptotically associated $d=1+2$
de Sitter spacetime.} this spread of f\/luctuations accounts for the reduction of the short distance scaling dimension,
e.g.\
instead of $d_{\rm sd}=2$ for the Proca f\/ield one arrives at $d_{\rm sd}=1$ for its stringlocal partner.
Whereas the $d_{\rm sd}$ for pointlike potentials increase with spin, their stringlike counterparts can always be
constructed in such a~way that their ef\/fective short distance dimension is the lowest one allowed~by
positivity, namely $d_{\rm sd}=1$ \textit{for all spins}.
It is not possible to construct the covariant ``inf\/inite spin''
f\/ields by the group theoretic intertwiner method used by Weinberg~\cite{Weinbook}; in~\cite{MSY,Pla} the more
powerful setting of modular localization was used.
In this way also the higher spin string-localized f\/ields were constructed.

For f\/inite spins the unique Wigner representation always has many covariant pointlike rea\-li\-za\-tions;
the associated quantum f\/ields def\/ine
linear covariant generators of the system of localized operator algebras whereas their Wick powers are nonlinear composite f\/ields.
In the following we will explain the reasons why even in case of pointlike generation one is interested in stringlike generating
f\/ields~\cite{MSY}.

For pointlike generating covariant f\/ields $\Psi^{(A,\dot{B})}(x)$ one f\/inds the following possibilities which link the
physical spin~$s$ to the (undotted, dotted) spinorial indices
\begin{gather}
\vert A-\dot{B} \vert \leqslant s\leqslant A+\dot{B}, \qquad m>0,
\label{spin}
\\
  h=A-\dot{B}, \qquad m=0.
\label{res}
\end{gather}
In the massive case all possibilities for the angular decomposition of two spinorial indices are allowed, whereas in the
massless case the values of the helicities~$h$ are severely restricted \eqref{res}.
For ($m=0$, $h=1$)
the formula conveys the impossibility of reconciling pointlike vector potentials with the Hilbert space positivity.
This clash holds for all $(m=0$, $s\geq0):$ pointlike localized ``f\/ield strengths'' (for $h=2$, the linearized Riemann
tensor) have no pointlike quantum ``potentials'' (got $h=2$, the $g_{\mu\nu}$, \dots) and similar statement holds for
half-integer spins in case of $s>1/2$. Allowing stringlike generators the possibilities of massless spinoral $A$, $\dot{B}$
realizations cover the same range as those in~\eqref{spin}.

Since the classical theory does not care about positivity, (Lagrangian) quantization does not guaranty that the expected quantum objects are consistent with the Hilbert space positivity; in fact it is well-known that
the gauge theoretic description necessitates the use of indef\/inite metric Krein spaces (the Gupta--Bleuler or BRST formalism).
The intrinsic Wigner representation-theoretical approach
 on the other hand keeps the Hilbert space and lifts the restriction to
pointlike generators in favor of semiinf\/inite stringlike generating f\/ields.

It is worthwhile to point out that contrary to popular belief perturbation theory does not require the validity of Lagrangian/functional
quantization.
Euler--Lagrange quantization limits the covariant realizations of $(m,s)$ Wigner representations to
a~few spinorial/tensorial f\/ields with low $(A,\dot{B})$ but as Weinberg already emphasized for setting up perturbation
theory one does not need Euler--Lagrange equations to formulate Feynman rules; they are only necessary if one uses formulation in which the
interaction-free part of the Lagrangian enters as it does in the Lagrangian/functional quantization.
The only ``classical'' input into causal perturbation as the E-G approach is a~(Wick-ordered) Lorentz-invariant f\/ield
polynomial which implements
the classical pointlike coupling, all subsequent inductive steps use quantum causality~\cite{E-G}.

\subsection{Modular localization and stringlocal quantum f\/ields}

An abstract modular~$S$-operator is a~closed antilinear involutive operator in Hilbert space~$H$ with a~dense domain of
def\/inition
\begin{gather*}
 {\rm Def.}~S:~{\rm antilin,~densely~def.,~closed,~involutive}~S^{2}\subseteq \mathbf{1},
\\
{\rm polar~decomp.}~S=J\Delta^{1/2},
\qquad
J~\text{modular~ref\/lection},
\qquad
\Delta^{it}
\
\text{mod.~group}.
\end{gather*}
Such operators have been f\/irst introduced in the context of the Tomita--Takesaki theory of (von Neumann) operator
algebras and are therefore referred to as ``Tomita~$S$-operator'' within the setting of operator algebras $\mathcal{A}$
by Tomita and Takesaki
\begin{gather*}
SA\Omega=A^{\ast}\Omega,
\qquad
A\in\mathcal{A},
\qquad
\text{action~of}~\mathcal{A}~\text{on}~\Omega~\text{is~standard},
\\
S=J\Delta^{1/2},
\qquad
J~\text{modular~ref\/lection},
\qquad
\Delta^{i\tau}=e^{-i\tau H_{\rm mod}}~\text{mod.~group}.
\end{gather*}
Here \textit{standardness} of the pair ($\mathcal{A}$, $\Omega$) means that the action is cyclic,
i.e.~$\overline{\mathcal{A}\Omega}=H$ and separating, i.e.~$A\Omega=0$, $A\in\mathcal{A}$ implies $A=0$, where the
separating property is needed for the uniqueness of~$S$. In quantum physics one meets such operators in equilibrium
statistical mechanics and QFT.
According to the Reeh--Schlieder theorem each local subalgebra $\mathcal{A}(\mathcal{O})$ is standard with respect to the
vacuum~$\Omega$ (in fact with respect to every f\/inite energy state)~\cite{Haag}.
In case of the wedge region $\mathcal{O}=W$, the operators which appear in the polar decomposition are well known in
QFT: $J$~is the ref\/lection along the edge of the wedge (the TCP operator up to a~$\pi$-rotation within the edge of the
wedge) whereas $\Delta^{i\tau}=U(\Lambda_{W}(\chi=-2\pi \tau))$ is the unitary representation of the~$W$-preserving
one-parametric Lorentz-boost group.

The modular localization theory has an interesting application within Wigner's po\-si\-ti\-ve energy representations
of the connected (proper, orthochronous) part of the Poincar\'{e} group~$\mathcal{P}_{+}^{\uparrow}$ as explained in the
following.
It has been realized, f\/irst in a~special case~\cite{AOP}, and then in the general setting~\cite{BGL} (see
also~\cite{Fa-Sc,MSY}), that there exists a~\textit{natural localization structure} on the Wigner representation
space for any positive energy representation of the proper Poincar\'{e} group.

Let $W_{0}$ be a~reference wedge region $W_{0}= \{z> \vert
t \vert;\, \mathbf{x}=(x,y)\in\mathbb{R}^{2} \}$. Such a~region is naturally related with two commuting
transformations: the $W_{0}$-preserving Lorentz-boost subgroup $\Lambda_{W_{0}}(\chi)$ and the $\mathbf{x}$-preserving
ref\/lection on the edge of the wedge $r_{W_{0}}$ which maps the wedge into its causal complement $W_{0}^{\prime}$. The
product of $r_{W_{0}}$ with the total ref\/lection $x\rightarrow-x$ is a~transformation in
$\mathcal{P}_{+}^{\uparrow}$, namely a~$\pi$-rotation in~$x$-$y$ edge.
On the other hand the total ref\/lection is the famous TCP transformation which, in order to preserve the energy
positivity, has to be represented by an anti-unitary operator.
With the only exception of zero mass f\/inite helicity representations where one needs a~helicity doubling (well known
from the photon representation), the total ref\/lection and hence $r_{W_{0}}$ is anti-unitarily represented on the
irreducible Wigner representation.
The resul\-ting operator\footnote{We keep the same notation as in the Tomita--Takesaki operator setting since the
dif\/ference between the algebraic and the representation theoretic~$S$ is always clear from the context.} is $J_{W_{0}}$
together with the commuting Lorentz boost $\Delta_{W_{0}}^{i\tau}$. Its analytic continuation $\Delta_{W_{0}}^{z}$ is an
unbounded operator whose dense domain in the one-particle space decreases with increasing $ \vert
\operatorname{Re}z \vert$. The anti-unitarity of~$J_{W_{0}}$ converts the commutativity with $\Delta^{i\tau}$ into the
relation $J_{W_{0}}\Delta_{W_{0}}^{a}=\Delta_{W_{0}}^{-a}J_{W_{0}}$ on a~dense set with the result that
\begin{gather*}
S_{W_{0}}=J_{W_{0}}\Delta_{W_{0}}^{1/2}, \qquad S_{W_{0}}^{2}\subset 1, \qquad {\rm i.e.} \quad {\rm Range}(S_{W_{0}})={\rm Dom}(S_{W_{0}})
\end{gather*}
is the polar decomposition of a~Tomita~$S$-operator.

With a~general $W$ def\/ined by covariance $W=gW_{0}$, where~$g$ is def\/ined up to Poincar\'{e} transformations which leave
$W_{0}$ invariant, we def\/ine
\begin{gather*}
\Delta_{W}^{i\tau}=g\Delta_{W_{0}}^{i\tau}g^{-1},
\end{gather*}

Involutivity implies that the~$S$-operator has $\pm1$ eigenspaces; since it is antilinear, the $+$~space multiplied with~$i$
changes the sign and becomes the $-$~space; hence it suf\/f\/ices to introduce a~notation for just one real eigenspace
\begin{gather*}
K(W)=\big\{\text{domain~of}~\Delta_{W}^{\frac{1}{2}}, \;  S_{W}\psi=\psi\big\},
\\
J_{W}K(W)=K(W^{\prime})=K(W)^{\prime},
\qquad
\text{duality},
\\
\overline{K(W)+iK(W)}=H_{1},
\qquad
K(W)\cap iK(W)=0.
\end{gather*}
It is important to be aware that one is dealing here with \textit{real} (closed) subspaces~$K$ of the complex
one-particle Wigner representation space $H_{1}$.
An alternative is to directly work with the complex dense subspaces $K(W)+iK(W)$ as in the third line.
Introducing the \textit{graph norm} in terms of the positive operator~$\Delta$, the dense complex subspace becomes
a~Hilbert space $H_{1,\Delta}$ in its own right.
The upper dash on regions denotes the causal disjoint (the opposite wedge), whereas the dash on real subspaces means the
symplectic complement with respect to the symplectic form $\operatorname{Im}(\cdot,\cdot)$ on~$H$. All the def\/initions work for
arbitrary positive energy representations of the Poincar\'{e} group~\cite{BGL}.

The two properties in the third line are the def\/ining relations of what is called the \textit{standardness property} of
a~real subspace\footnote{According to the Reeh--Schlieder theorem a~local algebra $\mathcal{A}(\mathcal{O})$ in QFT is in standard
position with respect to the vacuum, i.e.~it acts on the vacuum in a~cyclic and separating manner.
The spatial standardness, which follows directly from Wigner representation theory, is just the one-particle projection
of the Reeh--Schlieder property.}; any abstract standard subspace~$K$ of an arbitrary real Hilbert space permits to def\/ine an abstract $S$-operator in its complexif\/ied Hilbert space
\begin{gather}
  S(\psi+i\varphi)=\psi-i\varphi, \qquad S=J\Delta^{\frac{1}{2}},
\label{inv}
\\
  \operatorname{dom} S=\operatorname{dom} \Delta^{\frac{1}{2}}=K+iK,
\nonumber
\end{gather}
whose polar decomposition (written in the second line) yields two modular objects, a~unitary modular group~$\Delta^{it}$
and an antiunitary ref\/lection which generally have however no geometric interpretation in terms of localization.
The domain of the Tomita~$S$-operator is the same as the domain of~$\Delta^{\frac{1}{2}}$, namely the real sum of
the~$K$ space and its imaginary multiple.
Note that for the physical case at hand, this domain is intrinsically determined solely in terms of the Wigner group
representation theory, showing the close relation between localization and covariance.

The~$K$-spaces are the real parts of these complex $\operatorname{dom} S$, and in contrast to the complex domain spaces they are closed
as real subspaces of the Hilbert space (corresponding to the one-particle projection of the real subspaces generated by
Hermitian f\/ield operators). Their symplectic complement can be written in terms of the action of the
$J$ operator and leads to the $K$-space of the causal disjoint wedge $W^{\prime}$ (Haag duality)
\begin{gather*}
K_{W}^{\prime}:=\{\chi\, |\, \operatorname{Im}(\chi,\varphi)=0,\; \text{all}\ \varphi\in K_{W} \}=J_{W}K_{W}=K_{W^{\prime}}.
\end{gather*}
The extension of W-localization to general convex causally complete spacetime regions $\mathcal{O=O}^{\prime\prime}$ is
done by representing the causally closed $\mathcal{O}$ as an intersection of wedges and def\/ining $K_{\mathcal{O}}$ as
the corresponding intersection of wedge spaces
\begin{gather}
K_{\mathcal{O}^{\prime\prime}}\equiv     {\displaystyle\bigcap\limits_{W\supset\mathcal{O}^{\prime\prime}}}
K_{W}, \qquad \mathcal{O}^{\prime\prime}=\text{causal~completion~of}~\mathcal{O}
\label{K}
\end{gather}
These $K$-spaces lead via~\eqref{inv} and~\eqref{K} to the modular operators associated with $K_{\mathcal{O}}$. For
arbitrary spacetime regions one def\/ines the $K$-spaces by ``exhaustion from the inside''
\begin{gather*}
K_{\mathcal{O}}=     {\displaystyle\bigcup\limits_{W\subset\mathcal{O}}}   K_{W}.
\end{gather*}
For irreducible Wigner representations the two spaces are equal but it is easy to construct QFTs in which this causal
completeness property is violated for simply connected convex region.
As explained before QFT models which violate causal completeness for simply connected convex spacetime regions $\mathcal{O}$,
i.e.~$\mathcal{A}(\mathcal{O})\varsubsetneq \mathcal{A}({\mathcal{O}}^{\prime\prime})$ are unphysical; this problem occurs in the context of
isomorphism between QFT in dif\/ferent spacetime dimensions (the AdS-CFT correspondence, next section).
It also limits Kaluza--Klein ideas of dimensional reductions to quasiclassical approximations.

Modular theory encodes localization properties of particle states into domain properties of Tomita $S$-operators;
re-expressing the $K$-space properties in terms of Tomita $S$-operators the causal disjoint property between regions
$\mathcal{O}_{1}\rangle \langle \mathcal{O}_{2}$ reads for integer spin representations~\cite{MSY}
\begin{gather}
S_{\mathcal{O}_{1}}\subset S_{\mathcal{O}_{2}}^{\ast}.
\label{S}
\end{gather}
Modular theory is the only known theory in which the operator content is fully encoded into domains.
Def\/ining f\/ield operators for $\mathcal{O}$-localized Wigner wave functions as
\begin{gather*}
\Phi(\psi)   =a^{\ast}(\psi)+a(S_{\mathcal{O}}\psi),
\qquad
S_{\mathcal{O}}\Phi(\psi)\Omega   =\Phi(\psi)^{\ast}\Omega.
\end{gather*}
$S$ acts as in the second line, independent of $\mathcal{O}$; the only dif\/ference is the
$\mathcal{O}$-dependent domain.
 The commutator of two $\Phi$ equals
\begin{gather*}
\left[\Phi(\psi_{1}),\Phi(\psi_{2})\right]   =(S_{\mathcal{O}_{1}}
\psi_{1},\psi_{2})-(S_{\mathcal{O}_{2}}\psi_{2},\psi_{1})
=0 \qquad \text{in~case~of~\eqref{S}}.
\end{gather*}
These operators implement a~functorial relation between (localized) Wigner $K$-subspaces and interaction-free (localized)
operator subalgebras of~$B(H)$ where~$H$ is the Hilbert space which is generated by the successive action of
$\Phi^{\prime}s$ on the vacuum $\Omega$. The functorial map is
\begin{gather*}
K_{\mathcal{O}}\rightarrow\mathcal{A}(\mathcal{O})=\text{Alg}\big\{e^{i(\Phi(\psi))}\big|\psi\in K_{\mathcal{O}}\big\}.
\end{gather*}
For half-integer (Fermion) representation there is a~corresponding graded functor.

In the presence of interactions this functorial relation between particle subspaces and localized algebras is lost.
What remains is a~rather weak relation between wedge-local particle states and their ``emulation'' in terms of applying
interacting operators af\/f\/iliated to a~wedge-local interacting algebra to the vacuum.

In order to make contact with the notion of generating covariant f\/ields one needs intertwiners which map covariant
$\mathcal{O}$-supported testfunctions into $\mathcal{O}$-localized Wigner wave functions.

For those who are familiar with Weinberg's intertwiner formalism~\cite{Weinbook} relating the $(m,s)$ Wigner
representation to the dotted/undotted spinor formalism, it may be helpful to recall the resulting ``master formula''
\begin{gather}
 \Psi ^{(A,\dot{B})}(x)=\frac{1}{\left( 2\pi \right) ^{\frac{3}{2}}}\int
\big(e^{-ipx}u^{(A,\dot{B})}(p)\cdot a(p)+e^{ipx}v^{(A,\dot{B})}(p)\cdot b^{\ast
}(p)\big)\frac{d^{3}p}{2\omega }, \label{field}   \\
 u^{(A,\dot{B})}(p)\cdot a(p):=
 \begin{cases}
\displaystyle \sum_{s_{3}=-s}^{s}u^{(A,\dot{B})}(p,s_{3})a(p,s_{3}), & m>0,\\
\displaystyle \sum_{h=\pm
\left\vert h\right\vert }u^{(A,\dot{B})}(p,h)a(p,h), & m=0,
\end{cases}  \nonumber
\end{gather}
where the $a$, $b$ creation operators correspond to the Wigner momentum space wave functions of particles/antiparticles and the
$u$, $v$ represent the intertwiner (between the unitary Wigner representation and its covariant description) and its charge conjugate. In other words
the covariant pointlike Wigner wave function becomes the positive
frequency part of the f\/ield operator, i.e.\ covariant wave functions and pointlike covariant f\/ields are functorially related.

For the third class (inf\/inite spin, last line), the sum over spin components has to be replaced by an inner product
between a~$p$, $e$-dependent inf\/inite component intertwiner~$u$ and an inf\/inite component $a(p)$ since in this case
Wigner's ``little space'' is inf\/inite-dimensional.
The $\Psi(x)$ respectively $\Psi(x,e)$ are ``generating wave functions'', i.e.~they are wavefunction-valued Schwartz
distributions which by smearing with $\mathcal{O}$-supported test functions become $\mathcal{O}$-localized wave
functions.
Adding the opposite frequency antiparticle contribution one obtains the above formula which, by re-interpreting the
$a^{\#}$, $b^{\#}$ as creation/annihilation operators (second quantization functor), describes point-respectively string-local free f\/ields.
The resulting operator-valued Schwartz distributions are ``global'' generators in the sense that they generate
$\mathcal{O}$-localized operators $\Psi(f)$ for \textit{all} $\mathcal{O}$ by ``smearing'' them with
$\mathcal{O}$-supported test functions.

Only in the massive case the full spectrum of spinorial indices $A$, $\dot{B}$ is exhausted~\eqref{spin}, whereas the
massless
situation leads to the restrictions~\eqref{res} which are the def\/ining property of (generalized) f\/ield strengths;
pointlike potentials violate the positivity of Wigner's representation
spaces.
This awareness about the possible conceptual clash between pointlike  localization and the Hilbert space is important for the introduction of
string-localization.

{\sloppy Whereas Weinberg \cite{Weinbook} constructs the $u,v$-intertwiner functions
of pointlike covariant f\/ields~\eqref{field} with the help of group theoretic
methods, the modular localization approach is based on the direct
construction of localized Wigner subspaces\footnote{Modular localized subspaces of positive energy Wigner representations were f\/irst
constructed in~\cite{BGL}. Before such concepts were used for the solution of integrable models~\cite{S1,AOP}.
The construction of stringlocal covariant f\/ields can be found in~\cite{MSY}.} and their generating
stringlocal wave functions and associated f\/ields. In that case the
intertwiners depend on the spacelike direction~$e$ which is not a parameter but, similar to the point~$x$, a variable in terms of which the
f\/ield f\/luctuates~\cite{MSY}. Its presence allows the short distance
f\/luctuations in~$x$ to be more temperate than in case of pointlike f\/ields.

}

The short-distance reducing property of the generating stringlike f\/ields is indispensable in the implementation of
renormalizable perturbation theory in Hilbert space for interactions involving spins $s\geq 1$ f\/ields\footnote{These are also
prcisely those interactions in which the absence of mass gaps does not lead to problems with the particle structure.}.
Whereas pointlike f\/ields are the mediators between classical and quantum localization, the stringlike
f\/ields are outside the Lagrangian or functional quantization setting since they are not solutions of Euler--Lagrange
equations; enforcing the latter one arrives at pointlike f\/ields in Krein space.
String-localization lowers the power-counting limit, but
requires a nontrivial extension~\cite{Jens,Schroer} of the iterative Epstein--Glaser renormalization machinery~\cite{E-G}.
In the next section it will be shown that modular localization is essential for \textit{generalizing Wigner's intrinsic
representation theoretical approach to the} (non-perturbative) \textit{realm of interacting localized observable algebras}.

In order to arrive at Haag's algebraic setting of local quantum physics in the absence of interactions one may avoid
``f\/ield coordinatizations'' and apply the Weyl functor~$\Gamma$ (or its fermionic counterpart) directly to \textit{wave
function subspaces} where upon they are functorially passing directly to operator algebras, symbolically indicated~by
the functorial relation
\begin{gather*}
K_{\mathcal{O}}\overset{\Gamma}{\rightarrow}\mathcal{A}(\mathcal{O}).
\end{gather*}
The functorial map $\Gamma$ \textit{also} relates the modular operators $S$, $J$, $\Delta$ from the Wigner wave function
setting directly with their ``second quantized'' counterparts $S_{\rm Fock}$, $J_{\rm Fock}$, $\Delta_{\rm Fock}$ in Wigner--Fock space; it
is then straightforward to check that they are precisely the modular operators of the Tomita--Takesaki modular theory
applied to causally localized operator algebras (using from now on the shorter~$S$,~$J$,~$\Delta$ notation for modular objects
in operator algebras)
\begin{gather*}
\sigma_{t}(\mathcal{A}(\mathcal{O}))   \equiv\Delta^{it}\mathcal{A}(\mathcal{O})\Delta^{-it}=\mathcal{A}(\mathcal{O}),
\\
J\mathcal{A}(\mathcal{O})J   =\mathcal{A}(\mathcal{O})^{\prime}=\mathcal{A}(\mathcal{O}^{\prime}).
\end{gather*}
In the absence of interactions these operator relation are consequences of the modular relations for Wigner
representations.
The \textit{Tomita--Takesaki theory secures their general existence for standard pairs $(A, \Omega)$}, i.e.~an
operator algebras $\mathcal{A}$ and a~state vector $\Omega\in H$ on which $\mathcal{A}$ acts cyclic and separating (no
annihilators of~$\Omega$ in~$\mathcal{A}$).
The polar decomposition of the antilinear closed Tomita~$S$-operator leads to the unitary modular automorphism group
$\Delta^{it}$ associated with the subalgebra $\mathcal{A}(\mathcal{O})\subset B(H)$ and the vacuum state vector~$\Omega$, i.e.~with
the pair $(\mathcal{A}(\mathcal{O}),\Omega)$.

Although $B(H)$ is generated from the two commuting algebras $\mathcal{A}(\mathcal{O})$ and $\mathcal{A}(\mathcal{O})^{\prime}$, they do not
form a~tensor product in $B(H)$. Hence the standard quantum-information concepts concerning entanglement and density
matrices are not applicable; the QFT realization of entanglement is stronger\footnote{The localization entropy of the
vacuum entanglement for $\mathcal{A}(\mathcal{O})/\mathcal{A}({\mathcal{O}}^\prime$) is inf\/inite.}
since the vacuum state (in fact any f\/inite energy state) restricted to a local
algebra~$\mathcal{A}(\mathcal{O})$ is an impure state which cannot be represented by a~density matrix. In fact the ``monad'' (the unique hyperf\/inite type
${\rm III}_1$ von Neumann factor algebra)~$\mathcal{A}(\mathcal{O})$ has neither pure states nor
density matrices.

Just in order to avoid confusions, modular localization of operators is more restrictive than modular localization of states.
A~state vector generated by
applying an algebraically indecomposable stringlike localized f\/ield to
the vacuum may have components to dif\/ferent irreducible pointlike
Wigner representations. Apart from the inf\/inite spin representations
the distinction between point- and string-like is limited to f\/ields and
has no relevance for particles. When one calls an electron an ``infraparticle'' one refers to the fact that its interacting physical f\/ield is
necessarily stringlocal and that its application to the vacuum, which
generates a highly reducible Wigner representation state, does not
contain a discrete component at $p^2 = m^2_e$.

The only case for which the modular localization theory (i.e.\ the adaptation of the Tomita--Takesaki modular theory to the
causal localization principle of QFT) has a~geometric interpretation (independent of whether interactions are present or
not and independent of the type of quantum matter), is the wedge region, i.e.~the Lorentz transforms of the standard
wedge $W= \{x_{0}<x_{3}\, |\, \mathbf{x}_{\rm tr}\in\mathbb{R}^{2} \}$. In that case the modular group is the
wedge-preserving Lorentz boost and the~$J$ represents a~ref\/lection on the edge of the wedge, i.e.~it is up to
a~$\pi$-rotation equal to the antiunitary TCP operator.
The derivation of the TCP invariance as derived by Jost~\cite{Jost}, together with scattering theory (the TCP
transformation of the $S$-matrix) leads to the relation
\begin{gather*}
J=S_{\rm scat}J_{\rm in},
\end{gather*}
which in~\cite{S1,AOP} has been applied to constructive problems of integrable QFTs.
This is a~relation which goes much beyond scattering theory; in fact it only holds in local quantum physics since it
attributes the new role of a~relative modular invariant of causal localization to the $S$-matrix which it does not have in~QM.

This opens an unexpected possibility of a~new access to QFT in which the f\/irst step is the construction of
generators for the wedge-localized algebra~$\mathcal{A}(W)$ with the aim to obtain spacelike cone-localized (with
strings as a~core) or double cone-localized algebras (with a~point as core) from intersecting wedge algebras.
In this top-to-bottom approach (which is based on the intuitive idea that the larger the localization region, the better
the chance to construct generators with milder vacuum polarization) f\/ields or compact localized operators would only
appear at the end.
In fact according to the underlying philosophy that all relevant physical data can be obtained from localized algebras
the use of individual operators within such an algebra may be avoided; the \textit{relative positioning of the localized
algebras should account for all physical phenomena in particle phenomena}.
The next section presents the f\/irst step in such a~construction.

The only prerequisite for the general (abstract) case is the ``standardness'' of the pair $(\mathcal{A},\Omega)$ where
``standard'' in the theory of operator algebras means that~$\Omega$ is a~cyclic and separating vector with respect to
$\mathcal{A}$, a~property \textit{which in QFT is always fulfilled} for localized $\mathcal{A}(\mathcal{O})$ (thanks to
the validity of the Reeh--Schlieder theorem~\cite{Haag}).
These local operator algebras of QFT are what I referred to in previous publications as
a~\textit{monad}~\cite{integrable}; their properties are remarkably dif\/ferent from the algebra of all bounded
operators~$B(H)$ which one encounters for Born-localized algebras in QM~\cite{interface}.
For general localization regions the one-parametric modular unitaries have no direct geometric interpretation
since they describe
a~kind of fuzzy algebraic automorphism (which only near the boundary inside~$\mathcal{O}$ permits a possible geometric visualization).
But they are uniquely determined in terms of modular intersections of their
geometric~$W$-counterparts and are expected to become important in any top-to-bottom construction of models of QFT.
Even in the simpler context of localized subspaces~$K_{\mathcal{O}}$ related to Wigner's positive energy
representation theory for the Poincar\'{e} group and its functorial relation to free f\/ields these concepts have shown to
be useful~\cite{BGL}.

The most important conceptual contribution of modular localization theory in the context of the present work is the
assertion that the reduction of the global vacuum (and more generally that of all physical (f\/inite energy)
states) to a~local operator algebra $\mathcal{A}(\mathcal{O})$ leads to a~thermal state for which the ``thermal Hamiltonian'' $H_{\rm mod}$ is the generator of the modular
unitary group
\begin{gather*}
e^{-i\tau H_{\rm mod}}:=\Delta^{i\tau},
\\
\langle AB \rangle = \langle Be^{-H_{\rm mod}}A \rangle,
\end{gather*}
where the second line has the form
of the KMS property known from
thermal states in the thermodynamic limit in which the Gibbs trace
formula for a box-enclosed systems passes to the GNS state state formulation for open systems~\cite{Haag}.
Whereas the trace formulation breaks down in the thermodynamic limit, this analytic KMS formula (asserting analyticity
in $-1<\operatorname{Im}\tau<0$) remains.
It is in this and only in this limit, that QM produces a~global monad algebra (dif\/ferent from~$B(H))$
whereas in QFT this situation
is generic since it arises for any $\mathcal{A}(\mathcal{O})$-restricted f\/inite energy state.

As mentioned in the introduction, the intrinsic thermal aspect of localization is the reason why the probability issue
in QFT is conceptually radically dif\/ferent from QM for which one has to add the Born probability requirement.

Closely related to a~modular localization is the ``GPS characterization'' of a~QFT (including its Poincar\'{e} spacetime
symmetry, as well as the internal symmetries of its quantum matter content) in terms of modular positioning of a~f\/inite
number of monads in a~shared Hilbert space.
For $d=1+1$ chiral models the minimal number of copies is 2, whereas in $d=1+3$ the smallest number for a~GPS construction
is 7~\cite{K-W}.
This way of looking at QFT is an extreme relational point of view in terms of objects which have no internal structure
by themselves; this explains the terminology ``monad'' (a realization of Leibnitz's point of view about reality in the
context of abstract quantum matter)~\cite{K-W,interface}.
As life is an holistic phenomenon, since it cannot be explained from its chemical ingredients, so is QFT, which cannot
be understood in terms of properties of a~monad.
This philosophical view of QFT exposes its radically holistic structure in the most forceful way; in praxis one
starts with one monad and assumes that one knows the action of the Poincar\'{e} group on it~\cite{S1,AOP}; this was
precisely the way in which the existence of factorizing models was shown~\cite{Lech}.

In order to show the power of this new viewpoint for ongoing experimentally accessible physics, the following last
subsection of this section presents some dif\/ferent viewpoint about some open problems in Standard Model physics.

\subsection{Stringlocal vector mesons and their local equivalence classes}

Modular localization theory shows how the conf\/lict between pointlike quantization and the Hilbert space positivity
structure, which appears in the Lagrangian or functional quantization of causally propagating classical f\/ield theory
involving higher spin $s\geq1$ f\/ields, can be avoided.
Instead of extending the quantum mechanical quantization rules to f\/ields, one should notice that the Hilbert space
\textit{quantum} causal locality places limits on the ``tightness'' of localization.
One may consider this as a~sharpening of the radical dif\/ferent nature of quantum f\/ields from their classical
counterparts; whereas the latter are simply ordinary functions, causal quantum f\/ields are operator-valued Schwartz
distributions.
It took a~long time to get used to the fact that quantum localization leads to the formation of singular vacuum
polarization clouds at the boundary of the localization region which only can be controlled by smearing pointlike f\/ields
with test functions with smoothly approach zero (surface roughening)\footnote{In the algebraic formulation in terms of
localized operator algebras of bounded operators this leads to the area proportional \textit{localization entropy}
(which diverges in the limit of vanishing roughness).}.
In fact before these aspects were understood, QFT was even suspected to be inconsistent as a~result of its ``ultraviolet
catastrophe''.

The general principles of QFT lead to an interesting connection between the mass gap hypothesis and localization.
QFT with mass gaps are generated by operators which are localized in arbitrarily narrow spacelike cones, i.e.~regions
whose core is a~semiinf\/inite spacelike string~\eqref{string}.
This theorem does not say anything about in which case mass gap QFT stringlocal generating f\/ields are really needed
in order to describe the physical content of the model, but at least one knows that one does not need operators
localized on higher-dimensional spacelike subregions.
In theories without a~mass gap as QED one knows that operators carrying a~Maxwell charge cannot be generated by the
pointlike f\/ields used in QED.
On the other hand massive QED can be formulated as an interaction between a~pointlike $s=1$ Proca f\/ield coupled to charged
spinor or scalar matter; but it is well known that as a~consequence of $d_{\rm sd}(A_{\mu}^{\rm P})=2$ instead of $1$ it turns
out to be non-renormalizable.
Using the fact that the short distance behavior can be improved by changing the problem to one in indef\/inite metric
(Krein) space and setting up an elaborate ghost formalism, the BRST gauge formulation can be formulated as pointlike
interaction in Krein space.

This poses the question whether the non-renormalizability of the Hilbert space matter f\/ields is the way in which the
model indicates that its full content cannot be described in terms of pointlike f\/ields so that it illustrates
a~non-trivial realization of the above theorem. In the following we will show that this is indeed the case, i.e.~behind the
BRST gauge formalism there exists a~stringlocal physical theory whose pointlike observables agree with the gauge
invariant operators of the BRST gauge description.
Pointlike physical matter f\/ields still exist but only in the form of very singular (nonrenormalizable) Jaf\/fe f\/ields with
a~very restricted testfunction smearing (and a~questionable role as generating f\/ields for localized operator
algebras)~\cite{Jaffe}.
Whereas the localization problem in $s\geq1$ zero mass interaction of the Wigner creation/annihilation operators~$a^{\#}(p.s_{3})$ already shows up in the nonexistence of pointlike
interaction-free potentials, its appearance in the case of interacting massive higher spin f\/ield interactions is more
discreet and happens through the connection of localization with renormalizability.

The unsolved problems which one encounters in trying to pass to physical operators in a~gauge theory formulation are
well known.
Formal expressions for physical matter f\/ields as stringlike composites in terms of gauge dependent pointlike f\/ields in
Krein space
\begin{gather}
\varphi(x,e)=\varphi^{\rm K}(x)e^{ig\int_{x}^{\infty}A_{\mu}^{\rm K}(x+\lambda e)e^{\mu}d\lambda}, \qquad e^{\mu}e_{\mu}=-1
\label{Jo}
\end{gather}
appeared already in publications of Jordan and Dirac during the 30s.
But anybody who, apart from playing formal games, tried to obtain a~computational control of such composite stringlocal
expressions, knows that this is an impossible task.
The new SLF setting inverts this problem from its head to its feet; instead of trying to represent physical
charge-carrying f\/ields in terms of pointlike f\/ields, it bases renormalized perturbation theory direct on stringlocal
f\/ields.
In this way one overcomes the clash between pointlike f\/ield in Krein space with the Hilbert space
structure~\cite{S1, charge}.

Although modular localization plays an important conceptual role in the physical resolution of the clash between
pointlike localization and the Hilbert space, most of the model calculations within the SLF setting can be done
directly.
An important role is played by the fact that pointlike massive free f\/ields and their stringlike siblings are linearly
related members of the same local equivalence class.
For $s=1$ the pointlike Proca f\/ield
\begin{gather*}
\left\langle A_{\mu}^{\rm P}(x)A_{\nu}^{\rm P}(x^{\prime})\right\rangle   =\frac {1}{(2\pi)^{3/2}}\int
e^{-ipx}M_{\mu\nu}^{\rm P}(p)\frac{d^{3}p}{2p_{0}},
\qquad
M_{\mu\nu}^{\rm P}(p)   =-g_{\mu\nu}+\frac{p_{\mu}p_{\nu}}{m^{2}}
\end{gather*}
is related to its stringlike counterpart as
\begin{gather}
A_{\mu}(x,e)   =A_{\mu}^{\rm P}(x)+\partial_{\mu}\phi(x,e),
\label{rel1}
\qquad
e^{\mu}A_{\mu}   =0, \qquad \partial^{\mu}A_{\mu}=-m^{2}\phi.
\end{gather}
This relation is a~direct consequence of the def\/inition of $A_{\mu}(x,e)$ and $\phi(x,e)$
\begin{gather}
F_{\mu\nu}(x)  :=\partial_{\mu}A_{\nu}^{\rm P}(x)-\partial_{\nu}A_{\mu}^{\rm P}(x), \qquad \partial^{\mu}F_{\mu\nu}=m^{2}A_{\mu}^{\rm P},
\label{rel2}
\\
A_{\mu}(x,e)  :=\int_{0}^{\infty}F_{\mu\nu}(x+\lambda e)e^{\nu}
d\lambda, \qquad \phi(x,e):=\int_{0}^{\infty}A_{\mu}^{\rm P}(x+\lambda e)e^{\mu} d\lambda.
\nonumber
\end{gather}
All three f\/ields are linear combinations of the same $s=1$ Wigner creation/annihilation opera\-tors~$a_{s_{2}}^{\#}(p)$, $s_{3}=-1,0,+1$ with~dif\/ferent linearly related intertwiner functions and the relation~\eqref{rel1}
follows from~\eqref{rel2} and $e^{\mu}A_{\mu}(x,e)=0$.\footnote{The massless stringlocal f\/ields look like
vector potentials in the axial gauge apart from a~signif\/icant conceptual dif\/ference: what led to their rejection in gauge
theory (the f\/ixed $e$ and the incurable singularity at $pe=0$), is here encoded into the essential SLF property:
variable directional f\/luctuations (distributions in~$e$).} The scalar stringlocal f\/ield $\phi(x,e)$ will be referred
to as the St\"{u}ckelberg f\/ield; but in contrast to the Krein space St\"{u}ckelberg f\/ield of the BRST gauge setting (see later)~$\phi$
is physical in the sense of acting in Hilbert space.

It has been shown in~\cite{MSY} that \textit{massive} scalar stringlocal f\/ields can interpolate any integer spin; in the
present case it creates $s=1$ particles from the vacuum.
This is not possible for massless particles; they can only be described by tensor potentials.
In fact in the massless limit the linear relation to pointlike potentials is lost and the only surviving f\/ield is the
$A_{\mu}(x,e)$. This can be explicitly seen by looking at the 2-point functions of the above f\/ields
\begin{gather}
M_{\mu\nu}^{AA}(p;e,e^{\prime})   =-g_{\mu\nu}-\frac{p_{\mu}p_{\nu}(e\cdot e^{\prime})}{(p\cdot e-i\varepsilon)(p\cdot
e^{\prime}+i\varepsilon)} +\frac{p_{\mu}e_{\nu}}{p\cdot e-i\varepsilon}+\frac{p_{\nu}e_{\mu}^{\prime}}{p\cdot e^{\prime}+i\varepsilon},
\label{class1}
\\
M^{\phi\phi}(p;e,e^{\prime})   =\frac{1}{m^{2}}-\frac{e\cdot e^{\prime}}{(p\cdot e-i\varepsilon)(p\cdot e^{\prime}+i\varepsilon)},
\qquad
M_{\mu}^{A\phi}=\dots, \qquad  \text{etc}.
\nonumber
\end{gather}
The appearance of mixed correlation between $A^{\rm P}$ and~$\phi$ is (dif\/ferent from BRST) due to the fact that the degrees
of freedom in terms of $a_{s_{2}}^{\#}(p)$ has been maintained.

The best way to interpret this situation is in terms of extending Borchers' concept of local equivalence classes of
f\/ields to stringlocal f\/ields.
The class of relative pointlike f\/ields (Wick polynomials in the case of free f\/ields) is known to play a~similar role as
coordinates in geometry: they describe the same model of~QFT.
The aforementioned theorem guaranties that this stays this way (at least in the presence of mass gaps) for local
equivalence classes of stringlocal f\/ields (with pointlike f\/ields being considered as $e$-independent f\/ields) since the
$S$-matrix is independent on what stringlocal f\/ield coordinatization from the local equivalence class one uses.
A~similar statement holds for half-integer spins.
The gain in using $A_{\mu}(x,e)$ instead of $A_{\mu}^{\rm P}(x)$ is the lowering of the short distance dimension
$d_{\rm sd}(A^{\rm P})=2\rightarrow d_{\rm sd}(A)=1$. The derivative of the St\"{u}ckelberg f\/ield compensates the leading short
distance term of~$A^{\rm P}$ at the price of string-localization: one unit of the nonrenormalizable interaction creating
pointlike~$A^{\rm P}$ has moved into directional~$e$-f\/luctuations leaving the minimal possible value $d=1$ for the end point
f\/luctuations in~$x$.

It is interesting to note that the local equivalence class picture permits a~generalization in which the linear relation
between $s=1$ free f\/ields is a~special case of a~more general relation for integer spin $s>1$ f\/ields
\begin{gather*}
A_{\mu_{1}\dots \mu_{n}}(x,e)=A_{\mu_{1}\dots \mu_{n}}^{\rm P}(x)+\partial_{\mu_{1}}
\phi_{\mu_{2}\dots \mu_{n}}+\partial_{\mu_{1}}\partial_{\mu_{2}}\phi_{\mu_{3}\dots \mu_{n}}+\cdots +\partial_{\mu_{1}}\cdots
\partial_{\mu_{n_{n}}}\phi.
\end{gather*}
The left-hand side represents a stringlocal $s=n$ tensor potential associated to a~pointlike tensor potential with
the same spin.
The $\phi^{\prime}s$ $s=n-i$, $i=1,\dots,n$ tensorial St\"{u}ckelberg f\/ields of dimension $d=n-i+1$.
Each~$\phi$ ``peels of\/f'' a~unit of dimension so that at the end one is left with the desired spin~$s$ stringlocal $d=1$
counterpart of the tensor potential analog of the Proca f\/ield.
The main problem of using such generalizations is the identif\/ication of those couplings which guaranty the existence of
suf\/f\/iciently many observables generated by pointlike Wightman f\/ields (operator-valued Schwartz distributions).
This may be important in attempts to generalize the idea of gauge theories in terms of SLF couplings involving massive
$s>1$ f\/ields.
It turns out that the stringlocal~$\phi$ f\/ields are inexorable ``escorts'' of stringlocal spin~$s$ tensor f\/ields.
Their presence in the interaction is the price to pay for converting nonrenormalizable spin s tensor f\/ields of dimension
$d=s+1$ into their stringlocal $d=1$ counterparts.
This will be explicitly illustrated for $s=1$ in the last subsection of this section.

The idea of SLF consists in starting from local zero-order equivalence class relations relations as~\eqref{rel1} and
show that they are either maintained in every order perturbation theory or replaced by other coupling dependent
relations.
In the case of massive QED the result is that~\eqref{rel1} can be maintained in every order (plausible since the
arguments in~\eqref{rel2} survive perturbation theory) but have to be complemented by an equivalence class relation for
the~$g$-coupled matter f\/ields
\begin{gather*}
\psi(x,e)=e^{ig\phi(x,e)}\psi(x).
\end{gather*}
In the present work we will be satisf\/ied with the simpler established property that the $S$-matrix is~$e$-independent,
i.e.~$S_{\rm scat}(e)=S_{\rm scat}^{\rm P}$. In that case the calculation only involves time-ordered products of free f\/ields (see
below).

The relations~\eqref{rel1} and~\eqref{rel2} have the appearance of gauge transformations in the BRST gauge setting.
But their conceptual content is quite dif\/ferent: instead of describing gauge transformations between pointlike gauge
f\/ields, their role in SLF is to \textit{relate string- with point-localized quantum fields within the same local
equivalence class.} Unlike the BRST gauge setting which maintains the quantization parallelism to classical gauge
theory, the SLF relations are simply consequences of the foundational modular localization property of QFT.

Before passing to the calculation of the second-order $S$-matrix, it is instructive to point at some formal similarities
with the BRST formalism.
In the Krein setting the relation correspon\-ding to~\eqref{rel1}
\begin{gather*}
\partial^{\mu}A_{\mu}^{\rm K}+m^{2}\phi^{\rm K}\sim0,
\end{gather*}
where the equivalence sign is meant to indicate that it cannot hold as operator relation on Krein space\footnote{This is analogous to the Gupta--Bleuler formalism in QED where relations between gauge dependent
operators hold only on subspaces.}, one expects to f\/ind it in the cohomological descent to the Hilbert space.
In fact in order to work with true operator relations one has to introduce in addition to the negative metric
St\"{u}ckelberg f\/ield two fermionic scalar ghost f\/ields~$u$ and $\hat{u}$. In that formulation the cohomological
equivalence relations are replaced by operator relation involving the nilpotent~$s$-operation, e.g.\
$s\hat {u}=-i(\partial^{\mu}A_{\mu}^{\rm K}+m^{2}\phi^{\rm K})=0$ on $H_{\rm phys}$. To tighten the formal similarity with the BRST
formalism, it is helpful to rewrite the relation~\eqref{rel2} in terms of a~dif\/ferential form calculus in which
$d_{e}$ acts on a~zero form
\begin{gather}
d_{e}(A_{\mu}(x,e)-\partial^{\mu}\phi(x,e))=0,
\label{op}
\end{gather}
which follows from~\eqref{rel1} and the~$e$-independence of the Proca f\/ield.
In contrast to the abstract algebraic~$s$-operation the SLF localization setting uses the dif\/ferential form calculus.

This dif\/ferential form calculus can be used in order to express the string independence of interactions.
Assume that we start from a~pointlike nonrenormalizable massive QED $j^{\mu}A_{\mu}^{\rm P}$ interaction.
Using the current conservation it is easy to convert this into a~renormalizable stringlike interaction
\begin{gather}
\mathcal{L}^{\rm P}
=j^{\mu}(x)A_{\mu}^{\rm P}(x)=j^{\mu}(x)A_{\mu}(x,e)-\partial_{\mu}V^{\mu}(x,e), \qquad V^{\mu}=j^{\mu}(x)\phi(x,e),
\label{proca}
\\
\text{or} \qquad d_{e}(\mathcal{L-}\partial_{\mu}V^{\mu}(x,e))=0, \qquad \mathcal{L}:=j^{\mu}(x)A_{\mu}(x,e).
\nonumber
\end{gather}
Here $\mathcal{L}$ is the renormalizable ($d_{\rm sd}(\mathcal{L})=4$) interaction density and the derivative part disposes
(peels of\/f) the $d_{\rm sd}=5$ contribution as a~boundary term at inf\/inity, so that the pointlike f\/irst order
$S=\lim\limits_{g(x)\rightarrow g}\mathcal{L}^{\rm P}(g)=g\int\mathcal{L}^{\rm P}(x)d^{4}x$ is the same as that of the stringlike
interaction.
We will say that the two interactions are asymptotically equivalent
\begin{gather}
\mathcal{L}^{\rm P}\overset{\rm AE}{\sim}\mathcal{L}.
\label{bra2}
\end{gather}
The problem of showing the~$e$-independence of the second-order renormalizable $S$-matrix def\/ined in terms of is
$\mathcal{L}$ is~$e$-independent is obviously a~renormalization problem since the treatment of the singularities in the
``pointlike time ordering'' $T(\mathcal{L}^{\rm P}(x)\mathcal{L}^{\rm P}(x^{\prime}))$ has to be def\/ined in such a~way that it is
AE equivalent its stringlike counterpart $T(\mathcal{L}(x,e)\mathcal{L}(x^{\prime},e^{\prime}))$.

\subsection{Second-order calculations for abelian vector meson interactions}\label{Section3.3}

The strategy of the implementation of adiabatic equivalence starts with the zero-order relation~\eqref{proca} which is
used in the Bogoliubov formula for the perturbative physical $S$-matrix and the physical f\/ields.
For massive QED the interaction density $\mathcal{L}$
\begin{gather*}
\mathcal{L}(x,f)=\int def(e)\mathcal{L}(x,e),
\qquad
\mathcal{L}(x,e)=A_{\mu}(x,e)j^{\mu}(x),
\\
\mathcal{L}\equiv\mathcal{L}(g,f)=\int g(x)\mathcal{L}(x,f)dx,
\\
\psi_{\rm int}(x,f)  :=\frac{\delta}{i\delta h(x)}S(\mathcal{L})^{-1} S(\mathcal{L}+h\psi)\big|_{h=0}, \qquad S(\mathcal{L})=Te^{i\int
g(x)\mathcal{L} (x,f)dx}
\end{gather*}
leads, according to the formal Bogoliubov prescription, to the perturbative $S$-matrix as well as to f\/ields indicated for
the simplest case in the second line for the interacting Dirac spinor; time-ordered products of interacting products
originate from higher functional derivatives\footnote{In order to include f\/ield strengths one needs another source term,
i.e. $S(L+h\psi+kF)$.}.
The physical $S$-matrix results from the Bogoliubov~$S$-functional in the adiabatic limit $g(x)\equiv1$. The existence of
this limit is only guarantied in the presence of mass gaps.
The interacting f\/ields $\psi_{\rm int}^{\rm phys}(x,f)$ also require this adiabatic limit; but as a~result of the appearance of
the inverse~$S$-functional, the requirement for their existence are less stringent.
They are localized in a~spacelike cone with apex~$x$ and require the same renormalization treatment as a~pointlike $d=1$
f\/ield.

The smearing function in the string direction can be f\/ixed.
The resulting physical $\psi(x,f)$-f\/ield depends nonlinearly on~$f$ and is localized in a~spacelike cone with apex
at~$x$\footnote{The apex is also the point which is relevant for the Epstein--Glaser distributional continuation.}.
The~$e$-independence of the scattering matrix is equivalent to~$f$-independence for $f^{\prime}s$ normalized to $\int
f=1$. The~$f$-independence of $S_{\rm scat}$ is expected since there exists a~structural theorem stating that the $S$-matrix
in models with mass gaps is independent of spacelike cone in which the interpolating operator (the operator used in the
LSZ large time scattering limit) was localized~\cite{Haag}.
The adaptation of the St\"{u}ckelberg--Bogoliubov--Epstein--Glaser (SBEP) iterative formalism~\cite{E-G} to stringlocal
f\/ields requires to include string-crossing in addition to point-crossing.
For the second-order calculation it is not necessary to study the full systematics of this new phenomenon which the reader
will f\/ind in a~forthcoming publication by Mund~\cite{Jens}.

In the following we will present two second-order $S$-matrix calculations in which the problem of point- and
string-crossings can be dealt with using pedestrian methods.
Both models describe couplings of massive vector mesons to scalar f\/ields; in the f\/irst case the matter f\/ield is complex
(``scalar massive QED'') whereas the second model describes a~coupling to a~Hermitian f\/ield.
Whereas the application of the new SLF Hilbert space setting to massive QED shows the expected induction of the second-order quadratic $A_{\mu}$ dependence from the model def\/ining f\/irst-order interaction, the induction\footnote{Dif\/ferent
from counterterms which come with new parameters, induced terms depend only on the model-def\/ining f\/irst-order couplings
and the masses of the participating free f\/ields.} of terms in the second-order Hermitian coupling comes with some
surprises.
In that case there is no correspondence to a~classical f\/ield theory since an interaction in the massless case does not
exist and a~massive vector meson-$H$ coupling has no classical guidance which is the best prerequisite for encountering
surprises.

The proof of~$e$-independece~\eqref{bra2} of the tree contribution in massive scalar QED~\eqref{S} involves
a~renormalization problem for the two-point function
\begin{gather}
\left\langle T\partial_{\mu}\varphi^{\ast}\partial_{\nu}^{\prime} \varphi^{\prime}\right\rangle
=\left\langle T_{0}\partial_{\mu}\varphi^{\ast}\partial_{\nu}^{\prime}\varphi^{\prime}\right\rangle +g_{\mu\nu}c\delta(x-x^{\prime}),
\qquad
\left\langle T_{0}\partial_{\mu}\varphi^{\ast}\partial_{\nu}^{\prime}\varphi^{\prime}\right\rangle
\equiv\partial_{\mu}\partial_{\nu}^{\prime}\left\langle T_{0}\varphi\varphi^{\prime}\right\rangle,
\label{T}
\end{gather}
where the $T_{0}$ denotes the usual free f\/ield propagator without derivatives,~$c$ is a~free renormalization parameter
and the upper dash is used to avoid writing $\varphi(x^{\prime},e^{\prime})$ in order to shorten the notation; this
notation will also be used in all subsequent equations.
As a~result of the two derivatives, the two-pointfunctions on the left hand side involves f\/ields of scaling dimension~2
and hence has a~scaling degree 4, which accounts for the presence of a~delta function renorma\-li\-za\-tion term in
$T\mathcal{LL}^{\prime}|_{\text{1-contr}}$.

We def\/ine the one-contraction component (tree-component) of the second-order ``anomaly''~$\mathfrak{A}$ as
\begin{gather}
-\mathfrak{A}_e \equiv d_{e}(T_{0}\mathcal{L} \mathcal{L}^{\prime}
-\partial_{\mu}T_{0}V^{\mu}\mathcal{L}^{\prime})_{1}= -d_e N+\partial^{\mu}N_\mu.
\label{bra4}
\end{gather}
It is a measure of the violation of string-independence of the formal
second-order extension of the f\/irst-order $e$-independence~\eqref{proca}. The
one-contraction component is a Wick-ordered  product of
4~f\/ields which determines the second-order $S$-matrix. If we are able
to represent it in the form of two normalization terms $N$, $N_\mu$ we could
absorb these two terms into a renormalization of the second-order
expression. In this case $N$ can be used to def\/ine a~renormalized second-order tree component
\begin{gather*}
T_{0}\mathcal{L} \mathcal{L}^{\prime}\big|_1\to T_{0}\mathcal{L} \mathcal{L}^{\prime}\big|_1+N
\end{gather*}
and the $N_\mu$ can be likewise absorbed into local change of the derivative
contribution. In case the dif\/ferentiation acts inside the time-ordering
one has
\begin{gather*}
d_{e}(T_{0}\mathcal{L} \mathcal{L}^{\prime}
-\partial_{\mu}T_{0}V^{\mu}\mathcal{L}^{\prime})=0.
\end{gather*}
Hence the delta contributions come from the singular part in
\begin{gather}
\partial_{\mu}T_{0}V^{\mu}\mathcal{L}^{\prime}|_{1}    =-i\big(\partial_{\mu
}\partial^{\mu} \langle T_{0}\varphi^{\ast}\varphi^{\prime} \rangle
\varphi\overleftrightarrow{\partial_{\nu}^{\prime}}\varphi^{\ast\prime}\phi
A^{\nu\prime}+\text{h.c.}\big)_{\text{sing}}+\text{reg.},\label{eq39a}\\
\partial_{\nu}\delta(x-x^{\prime})\varphi\varphi^{\ast\prime}\phi A^{\nu
\prime}    =\partial_{\nu}(\delta(x-x^{\prime})\varphi\varphi^{\ast\prime}\phi
A^{\nu\prime})-\delta(x-x^{\prime})\partial_{\nu}(\varphi\varphi^{\ast\prime
}\phi A^{\nu\prime}),\label{eq39b}
\end{gather}
where the delta in \eqref{eq39b} results from the action of the wave
operator on the~$T_{0}$ propagator. The~$\partial_{\nu}^{\prime}$ derivative acting on the delta function can be replaced by a~$-\partial_{\nu}$ derivative which in turn can be rewritten as in~\eqref{eq39b}. The f\/irst term in~\eqref{eq39b} leads to~$\partial_{\nu
}N^{\nu}$ in~(\ref{bra4}).

Adding up all terms in \eqref{eq39a}  which contribute to the second
term in~\eqref{eq39b} one f\/inds that the contributions which involve
derivatives~$\partial_{\nu}$ of the $\varphi$-f\/ields cancel,
so that only the term for which the derivative acts on~$\phi$
remains. Taking into account the relation~$d_{e}\partial_{\nu}\phi
=d_{e}A_{\nu}$ we obtain
\begin{gather*}
d_{e}\delta(x-x^{\prime})\varphi\varphi^{\ast}\partial_{\nu}\phi A^{\nu\prime
}=d_{e}\delta(x-x^{\prime})\varphi\varphi^{\ast}A_{\nu}A^{\nu\prime}.
\end{gather*}
The derivation of the $e^{\prime}$ contribution to the anomaly
\begin{gather*}
-\mathfrak{A}_{e^{\prime}}=d_{e^{\prime}}(T_{0}\mathcal{L} \mathcal{L}
^{\prime}-\partial_{\mu}T_{0}\mathcal{L}V^{\prime\mu})_{1}
\end{gather*}
follows the same steps, so that the f\/inal result (including the
correct numerical coef\/f\/icients) is
\begin{gather}
(d_{e}+d_{e^{\prime}})(T_{0}\mathcal{LL}^{\prime}+2\delta(x-x^{\prime}
)\varphi^{\ast}\varphi A_{\nu}A^{\prime\nu})_{1}=\text{derivative terms}.
\label{2nd}
\end{gather}
Since derivative terms do not contribute to the adiabatic limit, our
result implies that the second-order $S$-matrix is a string-independent global
observable, i.e.\ the f\/irst-order adiabatic equivalence~(\ref{bra2}) has been
extended to second order. The renormalized interaction is of the same form
(apart from the fact that vector potential at the same point have dif\/ferent
directional variables~$e$) as that obtained from imposing the gauge
formalism in the form
\[
\partial_{\mu}\ \rightarrow \ D_{\mu}=\partial_{\mu}-igA_{\mu}.
\]
In the present case the only requirement was the Hilbert space setting
(which for~$s\geq1$ implies a weakening of localization from point-
to string-like).

The result may be conveniently written in the form
\begin{gather*}
T_{0}\mathcal{LL}^{\prime}\rightarrow T\mathcal{LL}^{\prime},\qquad  \langle
T\partial_{\mu}\varphi^{\ast}\partial_{\nu}\varphi \rangle = \langle
T_{0}\partial_{\mu}\varphi^{\ast}\partial_{\nu}\varphi\rangle -g_{\mu
\nu}\delta(x-x^{\prime}),
\end{gather*}
i.e.\ the renormalization can be encoded into a change of the two
derivative propagator where all propagators with a~lesser number of
derivatives remain unchanged. In this form it holds for tree contributions in
any order.

Encouraged by this success one may ask the question whether this
formalism can be generalized to def\/ine string-independent interaction densities. This is indeed possible. The
respective relation is more restrictive and has the form~\cite{Schroer}
\begin{gather}
   d\big(T\mathcal{LL}^{\prime}-\partial^{\mu}TV_{\mu}\mathcal{L}^{\prime
}-\partial^{\prime\nu}T\mathcal{L}V_{\nu}^{\prime}+\partial^{\mu}
\partial^{\prime\nu}TV_{\mu}V_{\nu}^{\prime}\big)=0,\qquad d=d_{e}+d_{e^{\prime}},\nonumber\\
   (T\mathcal{LL}^{\prime})^{\rm P}\equiv T\mathcal{LL}^{\prime}-\partial^{\mu
}TV_{\mu}\mathcal{L}^{\prime}-\partial^{\prime\nu}T\mathcal{L}V_{\nu}^{\prime
}+\partial^{\mu}\partial^{\prime\nu}TV_{\mu}V_{\nu}^{\prime},\label{eq44}
\end{gather}
where  $T$ stands for a renormalized~$T_{0}$. A~renormalized pointlike second-order interaction density can then be def\/ined as
in~\eqref{eq44}.

The delta function term~\eqref{2nd} corresponds to the second-order quadratic gauge contribution which in classical
gauge theory is subsumed in the substitution of $\partial_{\mu}$ by $\partial_{\mu}\rightarrow
D_{\mu}=\partial_{\mu}-ieA_{\mu}$. In our operator setting the second-order contribution~\eqref{2nd} has no intrinsic
signif\/icance since it is part of the renormalized $TLL^{\prime}$ which leads to the~$e$-independent $S$-matrix.
In fact the value of the normalization constant $c$ depends on the choice~$T_{0}$. The correct physical picture is that
of second-order ``induction'' from the f\/irst order; with other words certain counterterms, which in $s<1$ pointlike
setting would introduce new parameters, are f\/ixed by the causal localization principle which requires
the~$e$-independence of $S_{\rm scat}$.
This induction mechanism is of particular importance for the coupling to neutral matter to which we will move now.

Hermitian scalar f\/ields~$H$ coupled to massive vector meson (the charge-neutral counterpart of massive scalar QED) represents the Higgs f\/ield in the work of the BRST operator gauge treatment by the Z\"{u}rich group (see \cite{Scharf}
and references therein) and more recently in~\cite{Garcia}.
The terminology ``Higgs'' in their presentation had no relation to any symmetry breaking (the ``Higgs mechanism''); they
showed that the imposition of the gauge formalism on massive vector meson interactions with neutral particles
\textit{induces} second-order terms which together with the model-def\/ining f\/irst-order interaction takes the shape of
a~``Mexican hat'' potential.
There is some irony in this result because the Higgs mechanism had problems with the classical picture of gauge
variance; a~problem which was pointed out by several authors.
On the other hand the correct induction mechanism of the Mexican hat potential was solely based on the implementation of
operator BRST gauge formalism~\cite{Scharf}, or directly on the causal localization principle in the new Hilbert
space SLF setting.
It reveals that the Higgs mechanism is the result of a~formal manipulation
which has no relation to the correct interpretation in terms of a~renormalizable interaction of massive vector mesons
with Hermitian matter.


The formal similarity of the nilpotent $s$-operation of the BRST gauge
formalism with the $d_{e}$ dif\/ferential form calculus of the SLF Hilbert
space approach extends to higher order calculations. The one-contraction
component of the second order anomaly in terms of the nilpotent BRST $s$-operation reads~\cite{Scharf}
\begin{gather*}
-\mathfrak{A}_{s}=\big(sT_{0}\mathcal{LL}^{\prime }-\partial ^{\mu }T_{0}Q_{\mu }
\mathcal{L}^{\prime }-\partial ^{^{\prime }\mu }T_{0}\mathcal{L}Q^{\prime
}\big)_{1}.
\end{gather*}
It is a measure of the violation of gauge invariance which results from the
use of the ``unrenormalized'' time ordering of the product of f\/irst-order
interactions and its computation is the f\/irst step in the construction of a
gauge invariant second order $S$-matrix. Since the action of~$s$ is always
def\/ined in a~Krein space, we may economize on notation and omit the
superscripts~$K$ on the operators. For the special case of vanishing
self\/interaction~$H^{3}$ ($c=0$) in $\mathcal{L}$ the anomaly turns out to be
of the form~\cite{Scharf}
\begin{gather*}
- \mathfrak{A}_{s}=sN+sR+\partial ^{\mu }N_{\mu }, \\
 N=i\delta A\cdot A\big(H^{2}+\phi ^{2}\big),\qquad R=i\delta \frac{m_{H}^{2}}{m^{2}}
\left(H^{2}\phi ^{2}-\frac{1}{4}\phi ^{4}\right).
\end{gather*}
The expression for $N_{\mu }$ has been omitted since it renormalizes $T_{0}Q_{\mu }L^{\prime }+T_{0}LQ_{\mu }^{\prime }$ but does not contribute
to the second order interaction density $T_{0}LL^{\prime }$.

The $N$-term can be absorbed into a redef\/inition of the time ordered product
\begin{gather*}
T\partial _{\mu }H(x)\partial _{\nu }^{\prime }H(x^{\prime })
 = T_{0}\partial _{\mu }H(x)\partial _{\nu }^{\prime }H(x^{\prime })+\alpha
g_{\mu \nu }\delta (x-x^{\prime }), \\
T\partial _{\mu }\phi (x)\partial _{\nu }^{\prime }\phi (x^{\prime })
 = T_{0}\partial _{\mu }\phi (x)\partial _{\nu }^{\prime }\phi (x^{\prime
})+\beta g_{\mu \nu }\delta (x-x^{\prime })
\end{gather*}
by appropriate choice of $\alpha$, $\beta$, so that the second order result
is of the form
\begin{gather*}
\frac{g^{2}}{2}(T\mathcal{LL}^{\prime }+R).
\end{gather*}
Scharf~\cite{Scharf} then shows that a nontrivial trilinear $H$ self\/interaction $c\neq 0$ would lead to a third order anomaly unless one also introduces a
quadrilinear self-coupling $c^{\prime }H^{4}$. Whereas $c\sim g$, the latter
is of second order $c^{\prime}\sim g^{2}$ (with well-def\/ined numerical
coef\/f\/icients). The resulting~$\hat{R}$ is a~$4^{\rm th}$ degree polynomial in~$H$,~$\phi$. The f\/inal step consists in def\/ining a ``potential''~$V$ by combining
the f\/irst- and second-order~$H$,~$\phi$ contributions from~$\mathcal{L}$ and~$\hat{R}$ to a ``Mexican hat'' potential
\begin{gather*}
V  = V_{1}+\frac{1}{2}V_{2}=g^{2}\frac{m_{H}}{8m^{2}}\left(H^{2}+\phi ^{2}+\frac{2m}{g}H\right)^{2}-\frac{m_{H}^{2}}{2}H^{2}, \\
S^{(2)}  = 1+gi\int {\mathcal L}_{\rm tot}d^{4}x+i^{2}\frac{g^{2}}{2}\int d^{4}x\int d^{4}x^{\prime }T{\mathcal L}{\mathcal L}^{\prime},\qquad g{\mathcal L}_{\rm tot}=g{\mathcal L}+V.
\end{gather*}
By construction the second order $S$-matrix is BRST gauge invariant $sS^{(2)}=0$, although the Mexican hat potential by itself is not.

Note that this way of writing $V$ only serves the purpose of facilitating a
formal comparison with the alleged symmetry-breaking potential in which the
two physical masses are replaced by the parameters of the ``Higgs mechanism''
(a coupling strength and a f\/ield shift). But the correct description is in
terms of a coupling of a massive vector meson to a massive Hermitian f\/ield.
In fact renormalized perturbation theory always starts with the
model-def\/ining free f\/ields (including their masses) and a f\/irst-order
polynomial interaction between these f\/ields.

This result implies a radical conceptual change as compared to the Higgs
mechanism. The Mexican hat potential is not the symmetry-breaking part of
the def\/ining $A$-$H$ interaction, but it is \textit{induced} in second order
from the def\/ining $A$-$H$ interaction. The most surprising aspect of this
induction mechanism is the fact that the coupling strengths of the most
general renorma\-li\-zable model of interaction of a massive vector meson~$A$
with a Hermitian f\/ield~$H$ (which includes~$H^{3}$,~$H^{4}$ self\/interactions)
is \textit{uniquely determined in terms of the~$A$-$H$
coupling~$g$}. This kind of induction of additional couplings from
BRST gauge invariance of the $S$-matrix continues to hold in the SLF Hilbert
space setting. It also shows that the second order induced~$H^{4}$
interaction is dif\/ferent from a possible higher order loop contributions
(``box-graphs'') of a~renormalization-caused self\/interaction as known from
scalar QED.

The implementation of the $e$-independence of $S$ in the Hilbert space
setting conf\/irms these f\/indings of gauge theory, but it also leads to the
appearance of additional terms which contain $A$-f\/ields and derivative of~$\phi$ which cannot be incorporated into a Mexican hat potential. This is a~consequence of the Hilbert space positivity which replaces the Krein space St\"{u}ckelberg~$\phi ^{K}$ of the above BRST gauge setting (re-introducing
the omitted~$K$) by a physical stringlocal scalar intrinsic escort f\/ield $\phi$. The Hilbert space approach has the additional advantage to relate
the physical particle states directly with Wigner creation/annihilation
operators whereas in the BRST gauge setting additional cohomological steps
are necessary in order to recover physical states.

This places the construction of string-independent $S$-matrix in the Hilbert
space setting on safer conceptual grounds than that of gauge invariant
$S$-matrices in the BRST setting. Whereas there is no doubt about the
correctness of the BRST description of \textit{local observables in terms of
gauge invariant pointlike fields}, it is less clear to what extend this can
be extended to particle states beyond the vacuum sector. It remains
questionable whether the proven f\/ield-particle relations (as, e.g., the LSZ
scattering theory on which the connection between f\/ields and the $S$-matrix is
based) can be upheld in a Krein space setting of BRST gauge theory. The
construction of $e$-independent $S$-matrices and a more detailed comparison
with its $sS=0$ characterization in the gauge setting will be the subject of
forthcoming work of a collaboration with Jens Mund.


This raises the question why the inconsistency of the Higgs mechanism was not seen during the 40 year of its existence.
Well, it was seen by some people.
There have been several attempts to point at its metaphoric aspects, but all of them were eventually lost in the
maelstrom of time.
Swieca together with Ezawa~\cite{S-E} proved that behind Goldstone's Lagrangian model observation there was a~structural
theorem\footnote{Based on the Jost--Lehmann--Dyson representation, which in turn was derived from locality in a~Hilbert
space setting.} which relates spontaneous symmetry breaking with the appearance of a~zero mass Boson (the Goldstone
Boson).
This is the mechanism by which conserved currents to nonexisting (long-distance-diverging) charges.
It is helpful to recall what was known at that time about conserved currents and charges in the form of a~schematic
table:
\begin{gather*}
  \text{screening:} \ Q=\int j_{0}(x)d^{3}x=0, \qquad \partial^{\mu}j_{\mu}=0,
\\
  \text{spont.~symm.-breaking:} \ \int j_{0}(x)d^{3}x=\infty,
\\
\text{symmetry:} \ \int j_{0}(x)d^{3}x=\text{f\/inite} \neq 0.
\end{gather*}
Of \looseness=-1 special interest for theories involving massive vector mesons is the screening of the Maxwell charge, since it is the
characterizing property of interacting massive vector mesons.
This goes back to a~conjecture by Schwinger and was later established as a~structural theorem by Swie\-ca~\cite{Swieca}.
Whereas in massive QED there exists besides the identically conserved Maxwell current also the standard charge counting
current whose global charge counts the dif\/ference of charge and anticharge, there is also a~\textit{charge neutral model
in which the Maxwell current $($leading to the screened Maxwell charge$)$} is the only current.
In the limit of vanishing vector meson mass, the chargeless model approaches a~free massless vector meson.
In the massive case the
screening appears already in zero-order since the Maxwell current~$j_{\mu
}^{M}(x)\sim m^{2}A_{\mu }^{P}$ and the charge of the identically
conserved Proca ``current'' vanishes. The screening theorem~\cite{Sw} can be
explicitly verif\/ied in low-order perturbation theory. It shows that the
Maxwell charge in the abelian Higgs model remains screened (zero) and does
not diverge as in case of spontaneous symmetry breaking.

Schwinger invented the two-dimensional \textit{Schwinger model} in order to illustrate charge scree\-ning in
a~mathematically controlled situation, and Lowenstein and Swieca presented its full solution~\cite{Lo-Sw}.
Some years later Swieca, after presenting a~general structural proof of the Goldstone conjecture~\cite{S-E}, succeeded
to prove a~theorem (the ``Schwinger--Swieca screening theorem''~\cite{Sw}) which showed that charge screening is
a~structural consequence of massive Maxwell charges (an identically conserved current associated to a~$F_{\mu\nu}$ f\/ield
strength).
In order to save the Higgs model from its incorrect interpretation in terms of the ``Higgs mechanism'' (mass creation
through spontaneous symmetry breaking) and to direct attention to the fact that the physical content of the Higgs model
is a~realization of Schwinger's screening with a~Hermitian instead of a~complex matter f\/ield, he referred to it in all his
publications the \textit{Schwinger Higgs screening}.
Here the name Higgs stands for the statement that Schwinger's screening can also be realized with neutral matter
(coupling of massive vector mesons with Hermitian f\/ields).
His attempts to direct people away from a~misunderstanding towards genuine intrinsic physical properties failed; his
ideas succumbed to the maelstrom of time.

\looseness=-1
This poses the question of how an era, which was as rich in innovative foundational ideas as the three decades after
wwII, could develop into situation in which a~misunderstanding on a~particular but important problem dominated particle
theory for more than 40 years.
The f\/inal answer will be left to historians, but for the author it seems that the unfortunate concurrence of two causes
contributed to the present situation.
On the scientif\/ic side it was the idea (coming from gauge theory) that massless $s=1$ interactions (QED, QCD) are simpler
than their massive counterparts.
The Higgs mechanism of mass creation through spontaneous symmetry breaking imposed on scalar QED is the result of that
Zeitgeist.
We know nowadays, last but not least through the new SLF Hilbert space setting applied to interacting vector mesons, that
the opposite situation prevails: models with mass gaps fall well within the standard particle-f\/ield setting, whereas the
dif\/f\/icult problems of gluon/quark conf\/inement and a~spacetime understanding of QED collision theory involving
electrically charged particles only occur in their massless limits.

\looseness=-1
Even in the case of Goldstone's spontaneous symmetry breaking the important point is not the `manipulation of
a~Lagrangian by applying a~constant shift in f\/ield space, but rather the existence of a~conserved current whose charge
diverges as a~result of its coupling to a~zero mass boson.
In this case one may still tolerate the shift in f\/ield space as an anthropomorphic presentation of an intrinsic structural
statement, namely the connection of the broken symmetry (its diverging charge which fails to generate the symmetry) with
the existence of a~Goldstone boson.
To extend such manipulation to unphysical (gauge-dependent) charged f\/ields (there are no pointlike physical charged f\/ields) in scalar QED is meaningless
since gauge ``symmetry'' is not a symmetry but a~pre\-scription which permits to extract certain physical observables from a Krein space
formalism.

The price to pay is that one completely misses the renormalizable interaction of a~massive vector meson with
a~charge-neutral (Hermitian) matter f\/ield which hides behind the ``symmetry breaking and mass creating Higgs mechanism''.
Whereas one may always add to interacting massive vector mesons additional couplings to Hermitian matter f\/ields, their presence is not
needed to generate or sustain the mass of vector mesons. There exists however no renormalization theory of massive vector mesons in
Hilbert space without the presence of intrinsic stringlocal scalar escort f\/ields~$\phi$. As in the Ginsberg--Landau or BCS theory of superconductivity where one does not have to introduce outside degrees of freedom in order to obtain short-ranged vector potentials, there is
presently no indication that interacting massive vector meons in QFT
require the presence of additional degrees of freedom in the form of~$H$ f\/ields.

This raises the question whether the observed LHC events could be bound states of the intrincic escort f\/ield $\phi$ f\/ields. This is supported by the observation that in interactions of stringlocal massive vector mesons with charge matter or among themselves the $\phi$ appear
in the interaction densities at the places where the $H$-terms in the BRST formalism appears. This should not be surprising since $H$
states are indistinguishable from massive gluonium states. Could a~Higgs model, after the loss of the foundational Higgs mechanism, be
a phenomenological description of $\phi$-states? The new Hilbert space description for $s \geq 1$ interactions requires a re-investigation of the
gauge-theoretic Krein space description. Particular attention is required with respect to results which were obtained on high-energy behavior from Feynman graphs since their use is limited to the pointlike Krein space formalism.

Recently several books by several authors (Unzicker, Lopez Corredoira, \dots) appeared in which the present situation in
science and in particular in particle physics was critically analyzed.
The line of attack is predominantly against ``Big Science'' as represented by CERN and their handling of the Higgs issue.
In spite of their sometimes polemic arguments these authors are far from being crackpots; one of them (Unzicker)
actually discloses a~considerable amount of insider knowledge covering the activities of CERN during the last 3 decades.
He accuses Big Science to use Peter Higgs, a~very modest and shy individual who f\/irst exemplif\/ied the Higgs mechanism,
to present the discovery of a~new scalar chargeless particle as the centuries greatest contribution to particle physics
in order to justify the enormous amount of resources and manpower of present High Energy Physics.

Whatever conceptual changes the Standard Model and in particular gauge theories will undergo in the years ahead, the
clarif\/ication of the ``Higgs mechanism'' in terms of a~new Hilbert space setting of $s\geq1$ renormalization theory will
certainly play an important role in the extension of QFT to interactions involving higher spin f\/ields.
According to the best of my knowledge it is the f\/irst time that ongoing foundational changes in local quantum physics
come into direct contact with observational aspects of Standard Model particle physics.
Hopefully this will also lead to a~reduction of the deep schism between the large community of users of QFT and a~small
group of researchers exploring its still largely unknown terrain which is the main concern of the present paper.

\subsection{Nonabelian couplings and the SLF view of conf\/inement}

\looseness=-1
In the previous section it was shown that the Higgs mechanism is the result of a~conceptual misunderstanding of QFT.
The physical content of the abelian Higgs model, which remains after removing the meaningless idea of a~spontaneous mass
creation by postulating a~Mexican hat potential, is that of a~renormalizable coupling of a~massive vector potential to
a~Hermitian scalar matter f\/ield.
The Mexican hat potential is not an input for a~spontaneous mass creation, but rather describes the terms which the
requirement of string independence of the $S$-matrix induces within the renormalizable SLF Hilbert space setting.
This result conf\/irms previous f\/indings within the BRST gauge setting (see a~recent review~\cite{Garcia}).
These results invalidate the claim that massive vector mesons owe their mass to a~Higgs breaking mechanism; instead they
lead to the presence of scalar intrinsic escort f\/ields~$\phi$ of massive vector mesons, which is a~new epiphenomenon of
the SLF Hilbert space setting for renormalizable interactions involving higher spin $s\geq1$ f\/ields.

There is presently no indications that the Hilbert space positivity and locality of f\/ields lead to further impositions
on the renormalizable f\/ield content beyond the existence of the intrinsic escorts, but it would certainly be helpful to
have a~better understanding of the more elaborate self-interactions in terms of explicit second-order
calculations~\cite{M-S}.
The BRST gauge formulation leads to restrictions on the form of vector meson self-couplings to that which is known from
classical gauge symmetry formulated in the mathematical setting of f\/ibre-bundles; this is hardly surprising since this ghost
formulation results from the adaptation of the classical gauge symmetry.
The new SLF Hilbert space formulation on the other hand has no connection to gauge ideas, it has only the
perturbative adaptation of the foundational locality principle (modular localization) as its disposal.
The reduction of an Ansatz of the most general self-interaction between massive vector mesons with equal masses to the Yang--Mills
form in which the self-couplings are connected by a~Lie-algebraic structure is not the result of the imposition of
a~symmetry but rather follows from the consistency of perturbative renormalization with locality and the Hilbert space
positivity for self-interactions between vector mesons.
This phenomenon has no counterpart for $s<1$ pointlike interactions and this explains why it hitherto remained
overlooked.
It is the nonabelian analog of the second-order induced $A\cdot A$ contribution in the previous subsection.

In more concrete terms, the SLF Hilbert space reformulation of a~pointlike self-interaction of vector mesons with
arbitrary real $f_{abc}$ couplings and identical masses should lead to the expected Lie-algebraic restriction of~$f$
\begin{gather*}
\mathcal{L}^{\rm P}   =\sum\limits_{abc}f_{abc}F_{a}^{\mu\nu}A_{b,\mu}^{\rm P}A_{c,\nu}^{\rm P},
\qquad
\mathcal{L}^{\rm P}   =\mathcal{L+\partial}^{\mu}V_{\mu}.
\end{gather*}
In other words the symmetric form, which in the standard gauge setting is the result of the dif\/ferential geometric
properties of gauge symmetry and which the operator BRST setting incorporates through its ghost charge formalism, should
follow solely from foundational Hilbert space setting of the causal localization principle of QFT.
The requirement that the general pointlocal form can be rewritten in the form of the second line, with $\mathcal{L}$
being the stringlocal counterpart of $\mathcal{L}^{\rm P}$, leads to the total antisymmetry of~$f$. Since derivative
terms do not contribute to the $S$-matrix\footnote{The perturbative $S$-matrix is obtained from $g$-integrated time-ordered
products of interaction densities in the adiabatic limit $g(x)\rightarrow g$. Boundary terms from derivative
contributions vanish in massive theories (Section~\ref{Section3.3}).}, the physical interpretation of that relation is that it
guaranties the~$e$-independence of the $S$-matrix.
The formulation of this requirement in second order imposes the desired Lie algebraic restrictions on~$f$~\cite{M-S}.

More explicitly one has
\begin{gather*}
\mathcal{L}   =\sum f_{abc}\left\{F_{a}^{\mu\nu}A_{b,\mu}A_{c,\nu}
+m^{2}A_{a,\mu}^{\rm P}A_{b}^{\mu}\phi^{c}\right\}, \qquad V_{\mu}=\sum f_{abc} F_{a}^{\mu\nu}(A_{b,\nu}+A_{b,v}^{\rm P})\phi^{c},
\\
  d_{e}(L-\partial^{\mu}V_{\mu})=0\qquad \text{if} \ f_{abc} \ \text{are~totally~antisymmetric}.
\end{gather*}
The validity of the Jacobi identity and hence their Lie-algebraic nature follows from the formulation of~$e$
independence in second order~\cite{M-S}.
This is similar to Scharf's use of gauge invariance~\cite{Scharf}, except that in the SLF Hilbert space setting there is
no reference to a~gauge symmetry.

In agreement with our underlying philosophy which emphasizes the physical simplicity of massive models as compared to
the incompletely understood subtleties of their massless counterpart, we consider the massless Yang--Mills models as limits of
massive Yang--Mills couplings.
In other words our viewpoint is opposite to that of the Higgs mechanism.
Interestingly the implementation of our strategy leads to the appearance of a~kind of substitute of the Higges f\/ield:
the intrinsic escort f\/ields $\phi_{a}$ which appear already in f\/irst-order perturbation.

The appearance of Lie-algebraic restrictions from the $s\geq1$ implementation of QFT principles should be seen in
connection with the Doplicher--Roberts result~\cite{Haag}.
The latter states that the superselection structure following from the locality property of local observables can be
encoded into a~f\/ield-net extension of the observable net which is symmetric under the local action of a~compact group.
In perturbation theory one has to impose the specif\/ic symmetry in terms of relations between in principle independent
renormalization terms.
In the case of the symmetric appearance of induced interactions for $s\geq1$ in our new setting there simply exists no
less symmetric theory on whose renormalization theory we can impose symmetry conditions.
What has been hitherto considered as a~consequence of imposing local gauge symmetry is really the result of the general
principles of QFT.
Whether this perturbative observation can be backed up by a~structural theorem, as it was possible in the
Doplicher--Roberts superselection theory, remains to be seen.

\looseness=-1
An important dif\/ference of the new setting compared to pointlike perturbation theory is that the connection between
of\/f-shell properties and high energy behavior on-shell restrictions in terms of Feynman diagrams break down.
The presence of stringlocal propagators and stringlocal vertices (from string-crossings) invalidate phenomenological
arguments in favor of Higgs particles based on high-energy improvements of perturbative on-shell unitarity.
The SLF Hilbert space formalism leads to a~very subtle connection between the bad of\/f-shell behavior and its on-shell
improvements.
The new perturbation theory with its string-crossings cannot be encoded into Feynman diagrams.

Coming to the relation with the LHC experiment one should point out that these results cannot distinguish between
a~``gluonium'' bound state of the intrinsic escort~$\phi$, and an added~$H$-coupling.
Unfortunately the impossibility of understanding bound states within perturbation theory impedes reliable quantitative
predictions, but this is not dif\/ferent from the description of hadrons in terms of bound states of quarks.
Since the induced couplings of the intrinsic escort~$\phi$ to the vector potentials is indistinguishable from
a~Higgs--Kibble coupling, the latter could be a~phenomenological description of the SLF Hilbert space situation.

\looseness=-1
An even more important problem for the future path of QFT and the Standard Model is the question whether it is possible
to show that conf\/inement has a~clear physical meaning in Yang--Mills theories, i.e.~that it can be derived on the basis of the
infrared behavior of massless limits in expectation values of stringlocal physical (Hilbert space\footnote{The
positivity of Hilbert space is expected to play an important role in order to obtain the physical infrared behavior of
stringlike gluons/quarks as opposed to that of their unphysical counterparts in the gauge setting.}) massive gluon/quark
f\/ields.
The new setting strongly suggests that conf\/inement means that correlation functions, which besides pointlike
observable composite f\/ields (gluonium, hadrons) also contain stringlocal gluon and quark f\/ields, should vanish.
The only exception should be spacelike separated~$q$-$\bar{q}$ pairs whose string direction is parallel to the direction
of their spacelike separation.

Free zero mass string f\/ields, with the exception of those belonging to the third Wigner positive energy class (massless
inf\/inite spin f\/ields), are reducible strings, i.e.~they can be written as semi-inf\/inite integrals over pointlike f\/ield
strengths.
This is certainly not the case for \textit{interacting} massless gluon f\/ields since the lowest pointlike composites are
of polynomial degree~4.
As the free inf\/inite spin strings~\cite{dark}, their are their noncompact localization is irreducible.
Both noncompact types of strings have problems with causality which forbids their emergence from collisions of ordinary
particles (i.e.~particles localized in compact regions).
Abelian zero mass theories are somewhere in the middle; the vector potential strings are reducible, but this is not the
case for the strings of charged matter.

In the case of interacting gluon strings one mechanism (perhaps the only one) of avoiding contradictions with
causality which arise from their appearance in collisions of compact matter is that correlations which, besides
pointlike composites (gluonium f\/ields) also contain gluon f\/ields, must vanish.
The SLF setting presents a~realistic scenario to check such a~situation because, dif\/ferent from the BRST gauge setting,
there is a~natural physical covariant stringlocal massive gluon f\/ield which for $m\rightarrow0$ passes to its massless
physical counterpart; so a~proof would consist in showing that a~partial resummation of the leading logarithmic
contributions to the infrared divergences leads to a~zero result after interchanging the limit with the summation
of the leading terms.
The infraparticle situation of charged particles is less radical, since in that case one expects the correlation
functions of physical stringlocal charged f\/ields to remain infrared-f\/inite and to represent the SLF counterpart
of~\eqref{Jo}.
The Yennie--Frautschi--Suura argument~\cite{YFS} (generalizing previous model calculations by Bloch and Nordsiek) is based
on the logarithmic divergencies in an infrared cut-of\/f parameter $\lambda\rightarrow0$ which appear in the mass shell
restriction of the stringlocal physical charged matter.
From low-order logarithmic divergencies one reads of\/f the systematics of the leading contributions from the higher terms
and f\/inds a~coupling-dependent power behavior $\lambda^{f(g)}$.\footnote{In the spacetime LSZ scattering setting of
infraparticles the mass shell singularities have been softened; so that it cannot be compensated by the large time wave
packet behavior.} One then concludes that the $\lambda\rightarrow0$ limes (the scattering amplitude for charged
particle with a~f\/ixed f\/inite number of outgoing photons) vanishes and that the perturbative logarithmic divergencies
only appeared because of the illegitimate inversion of the perturbative expansion with the $\lambda \rightarrow0$ limit.
The authors~\cite{YFS} then show that a~nontrivial scattering information resides in photon inclusive cross sections
rather than scattering amplitudes; such a~construction has no of\/f-shell counterpart.

\looseness=1
The SLF setting suggests an interesting improvement of the YFS argument which consists in replacing the ad hoc
noncovariant infrared regulator~$\lambda$ by the covariant physical vector meson mass~$m$. The limit should reproduce the
YFS result of vanishing scattering amplitudes for charged particle scattering with a~f\/inite number of outgoing photons.
Another a~possible proof which is of a~more structural kind is to f\/irst show that the infrared properties replace the mass-shell
pole by a~less singular cut.
The resulting milder singularity cannot compensate the dissipation of wave packets which enter the LSZ formalism; this
leads the vanishing of the $t\rightarrow\infty$ LSZ limit.

In the Yang--Mills case one expects that the of\/f-shell correlation of massive gluons, which for $m\rightarrow0$ are
logarithmically divergent, vanish after interchanging the massless limit with the summation of leading logarithmic
divergencies.
Correlations which only contain pointlike composites are expected to stay f\/inite in this limit.
This would resolve the causality violating emergence of noncompact localized objects from compact spacetime collision
regions.
In a certain sense the this causality problem which is avoided through conf\/inement, is opposite to
that of the irreducible free strings of the inf\/inite spin Wigner positive
energy representations class; in that case one expects that apart, from
its coupling to gravitation (any kind of positive energy matter couples to gravitation), noncompact matter cannot change into ordinary
matter; this kind of inertness makes such matter an excellent candidate
for dark matter~\cite{dark}.

The SLF setting also presents a~rigorous perturbative way to check the asymptotic freedom property based on the beta
function in well-def\/ined Callen--Symanzik equations for well-def\/ined correlation functions.
The existing derivation is only a~consistency argument\footnote{The coef\/f\/icients of the Callen--Symanzik equations are global
quantities and as such cannot be computed solely on the basis of the known perturbative short distance behavior.} and
not a~proof; it shows that the educated guess of a~massless Yang--Mills beta function is consistent with the computable short
distance behavior.
A~rigorous calculation would f\/irst establish the Callen--Symanzik equations for the renormalizable stringlocal massive Yang--Mills coupling
and then appeal to the mass-independence of the beta function.

It is appropriate to end this section with two remarks which relate the present results to other ideas which arose in
the history of particle physics.

The SLF Hilbert space approach is vaguely reminiscent of Mandelstam's idea to formulate QED solely in terms of f\/ield
strengths.
It turns out that precisely the directional f\/luctuation of the $x+\mathbb{R}_{+}e$ localized $A_{\mu}(x,e)$ in~$e$ (a~point in $d=1+2$ de Sitter spacetime) attenuate the strength of the~$x$-f\/luctuations and renders the interaction
renormalizable in the sense of power-counting.
The picture is that the nonvanishing commutators for string crossing are \textit{necessary for lowering the singularity
for coalescent}~$x$. Mandelstam's approach failed because in his setting it seems to be dif\/f\/icult to take care of this
advantage~\cite{Mandel,Mandel+}.
In both, the massless as well as the massive case, there always exists a~string-localized description in which
the~$e$-f\/luctations lower the strength of the~$x$-f\/luctuation of the pointlike description in such a~way that the
resulting short distance scale dimension is $d=1$, independent of spin.

\looseness=1
As mentioned before, there is also a formal relation to the ``axial gauge''.
Although it was seen that this gauge is formally compatible with a Hilbert
space structure, the interpretation of~$e$ as a f\/ixed gauge
parameter (not participating in Poincar\'{e} transformations) misses its role
in the formulation of stringlocal renormalizable interactions which avoids the
use of unphysical Krein spaces. The special status which gauge
theory attributes to interactions of vectormesons has been removed and
$s=1$ interactions have been united with $s<1$ interactions
under the common roof of the localization principle in a Hilbert space
setting. In this way SLF leads to a~democratization of low and high
spin QFT under the shared conceptual roof of its foundational quantum causal
(modular) localization principle. This ``democratization'' on the level of
f\/ields parallels that of particles in that the new setting also removes the
hierarchical role of Higgs particle (the ``God particle'') and re-establishes
``nuclear democracy'' between particles.

\looseness=1
Stringlike localization also appears in the axiomatic approach as the
tightest localization which can be generically derived from a theory with
local observables and a mass gap (the existence of pointlike generators is
viewed as a special case of stringlike localization). This was the result of a~structural theorem by Fredenhagen and Buchholz  in the 80s~\cite{Haag}.
It is natural to think of the strings of matter f\/ields in massive gauge theories (which unlike the vector meson strings
cannot be removed by dif\/ferentiations) as Buchholz--Fredenhagen~\cite{Bu-Fr} spacelike-cone-localized objects whose singular
generators are strings.
Their description remains somewhat abstract and does not reveal the connection of stringlike f\/ields with the
perturbative nonrenormalizability and the singular Jaf\/fe (in contrast to Wightman) type structure.
As a~curious historical side remark it should be added that it was the improvement of Swieca's screening theorem
by Buchholz and Fredenhagen which led to their derivation of string-localization from the mass gap assumption.

\section[Generators of wedge algebras, extension of Wigner representation theory\\ in the presence of interactions]{Generators of wedge algebras, extension of Wigner\\ representation theory in the presence of interactions}\label{Section4}

Theoretical physics is one of the few areas of human endeavor in which the identif\/ication of an error may be as
important as the discovery of a~new theory.
This is especially the case if the error is related to a~lack of understanding or a~misunderstanding of the causal
localization principle which is the basis of QFT.
The more remote the properties of interest are related to the def\/ining causal localization properties of QFT, the more
speculative becomes the research and the larger is the probability to run into misunderstandings.

A perfect illustration of this point is the \textit{on-shell approach to particle theory} in connection with the study
of the $S$-matrix and formfactors in the aftermath of the successful application of dispersion relations to high energy
nuclear reactions in the late 50s.
Leaning on this limited but important success particle theorists in the 60s begun to use analytic on-shell properties
for general on-shell constructions as the \textit{$S$-matrix bootstrap} and Mandelstam's subsequent attempts to use
crossing symmetric two variable spectral representations for the actual construction of high energy nuclear elastic
scattering amplitudes.

Whereas of\/f-shell analytic properties of correlation function were systematically analyzed in the pathbreaking work of
Bargmann, Hall and Wightman~\cite{St-Wi}, it was already clear at the time of the dispersion relations that on-shell
analytic properties are of a~dif\/ferent conceptual caliber.
The analytic properties coming from the causal structure of correlation function could not account for the analytic
on-shell properties.
In particular the foundational origin of the important particle crossing property, one of the most subtle particle-f\/ield
connections, remained outside the range of at that time known methods, apart from some very special cases which were
solved with the help of the (still) intricate mathematics of several complex variables~\cite{BEG}.
The unfavorable relation between mathematical ef\/fort and meager physical result led to an end of these attempts.

Only after the arrival of modular localization and its role in the construction of $d=1+1$ integrable models~\cite{AOP}
for the spacetime localization aspects of the Zamolodchikov algebra structure, the situation began to improve.
The crucial step was the realization that the $S$-matrix was not only an operator resulting from time dependent scattering
theory (which it is in every~QT), but also represented a~relative\footnote{It connects the modular wedge localization of
the incoming f\/ields with that of the interacting wedge-local algebra.} modular invariant of wedge-localized algebras.
This led to the idea that the analytic particle aspects of the crossing property could be a~consequence of the analytic
KMS identity for wedge-localized algebras (after rewriting its f\/ield content into particle properties).
The resulting derivation of the particle crossing relation from the same modular localization principle which solves the
E-J conundrum and explains the Unruh ef\/fect~\cite{Bu-So, Unruh} is somewhat surprising; this and the closely
related proposal for a~general on-shell construction~\cite{integrable} which extends the successful approach of
integrable models from the structure of their generators of wedge algebras~\cite{Lech} will be the theme of this
section.

In this way the original aim of Mandelstam's on-shell project for f\/inding a~route to particle theory which is dif\/ferent
to quantization and perturbation theory (and stays closer to directly observational accessible objects) will be
recovered, and the errors which led to the dual model and ST will be avoided.
The new on-shell project is a~``top-to-bottom'' approach in which the aims and concepts are laid down before their
mathematical and computational implementation starts.
This is opposite to the perturbative SLF setting which starts from interaction densi\-ties~$\mathcal{L}$ in terms of
(string- or point-local) free f\/ields and tries to construct a QFT of interacting f\/ields.
What binds them together in this paper is that both of them are realizations of the quantum causal localization
principle.
The on-shell approach starts from the algebraic structure of generators of the wedge algebra (the Zamolodchikov--Faddeev
algebra in the integrable case) and sharpens the localization by constructing compact localized algebras as
intersections of wedge algebras.
Point- or string-local generating f\/ields of such algebras only appear, if at all, only at the very end\footnote{In the
LQP formulation one does not need them since all physical informations can be directly derived from the net of local
algebras~\cite{Haag}.}.

In order to motivate the reader to enter a~journey which takes him far away from text-book QFT, it is helpful to start
with a~theorem which shows that the familiar particle-f\/ield relations breaks down in the presence of \textit{any}
interaction.
The following theorem shows that the separation between particles and \textit{interacting} localized f\/ields and their
algebras is very drastic indeed~\cite{integrable}:

\begin{Theorem}[Mund's algebraic extension~\cite{Mund2} of the J-S theorem~\cite{St-Wi}] A~Poincar\'{e}-covariant QFT in $d\geq1+2$
fulfilling the mass-gap hypothesis and containing $($a sufficiently large set of$)$ ``temperate'' wedge-like
localized vacuum polarization-free one-particle generators $($PFGs$)$ is unitarily equivalent to a~free field theory.
\end{Theorem}

\looseness=-1
It will be shown below that the requirement of temperateness of generators (Schwartz distributions,
equivalent to the existence of a~translation covariant domain~\cite{BBS}) is a~very strong restriction; it only allows
integrable models, and integrability in QFT can only be realized in $d=1+1$.
Note that Wightman f\/ields are assumed to be operator-valued temperate distributions.
Hence the theorem says that even in case of a~weak localization requirement as wedge-localization one cannot f\/ind
interacting operators with reasonable domain properties which, as in Wightman QFT, allow their subsequent application.
However any QFT permits wedge-localized nontemperate generators~\cite{BBS}.
The theorem has a~rich history which dates back to Furry and Oppenheimer's observation (shortly after Heisenberg's
discovery of localization-cause vacuum polarization) that Lagrangian interactions always lead to f\/ields which, if
applied to the vacuum, inevitably create a~particle-antiparticle polarization cloud in addition to the desired
one-particle state.

The only remaining possibility to maintain a~relation between a~polarization-free generator
(PFG) leading to a~pure one-particle state and a~localized operator (representing the f\/ield side) has to go through the
bottleneck of \textit{nontemperate PFG generators of wedge-localized algebras}; this is what remains
of the non-interacting particle-f\/ield relation in the presence of interactions.

For the on-shell construction one needs also a~relation between \textit{multiparticle states} and (na\-tu\-rally
nontemperate) operators af\/f\/iliated to wedge algebra.
The idea is to construct a~kind of \textit{``emulation'' of free incoming fields} (particles) restricted to
a~wedge regions \textit{inside the interacting wedge algebra} as \textit{a replacement for the nonexisting second
quantization functor}.
As the construction of one-particle PFGs, this is achieved with the help of modular localization theory.

The starting point is a~\textit{bijection} between wedge-localized incoming f\/ield operators representing the particle
aspects and interacting wedge-local operators.
This bijection is based on the equality of the dense subspace which these operators of the two dif\/ferent algebras
create from the vacuum.
Since the domain of the Tomita~$S$ operators for two algebras which share the same modular unitary $\Delta^{it}$ is the
same, a~vector $\eta\in \operatorname{dom} S\equiv \operatorname{dom} S_{\mathcal{A}(W)}=\operatorname{dom} \Delta^{\frac{1}{2}}$ is also in
$\operatorname{dom} S_{\mathcal{A}_{\rm in}(W)}=\Delta^{\frac{1}{2}}$ (in~\cite{BBS} it was used for one-particle states).
In more explicit notation, which emphasizes the bijective nature, one has
\begin{gather*}
A \vert 0 \rangle   =A_{\mathcal{A}(W)} \vert
0 \rangle, \qquad A\in\mathcal{A}_{\rm in}(W), \qquad A_{\mathcal{A}(W)} \in\mathcal{A}(W),
\\
S(A)_{\mathcal{A}(W)} \vert 0 \rangle   =(A_{\mathcal{A}( W)})^{\ast} \vert
0 \rangle =S_{\rm scat}A^{\ast}S_{\rm scat}^{-1} \vert 0 \rangle, \qquad S=S_{\rm scat}S_{\rm in},
\\
S_{\rm scat}A^{\ast}S_{\rm scat}^{-1}\in\mathcal{A}_{\rm out}(W).
\end{gather*}
Here~$A$ is either an operator from the wedge localized free f\/ield operator algebra $\mathcal{A}_{\rm in}(W)$ or an
(unbounded) operator af\/f\/iliated with this algebra (e.g.\
products of incoming free f\/ields~$A(f)$ smeared with~$f$, $\supp f\in W$);
$S$~denotes the Tomita operator of the
interacting algebra~$\mathcal{A}(W)$ whereas $S_{\rm in}$ denotes that associated with the interaction-free incoming
algebra.
Under the assumption that the dense set generated by the dual wedge algebra $\mathcal{A}(W)^{\prime}\left\vert
0\right\rangle$ is in the domain of def\/inition of the bijective def\/ined ``emulats'' (of the wedge-localized free f\/ield
operators inside its interacting counterpart) the $A_{\mathcal{A}(W)}$ are uniquely def\/ined; in order to be able to use
them for the reconstruction of~$\mathcal{A}(W)$ the domain should be a~core for the emulats.
Unlike smeared Wightman f\/ields, the emulats $A_{\mathcal{A}(W)}$ do not def\/ine a~polynomial algebra, since
their unique existence does not allow to impose additional properties; in fact they only form a~vector space and the
associated algebras have to be constructed by spectral theory or by other means to extract an algebra from a~vector
space of closed operators (as Connes reconstruction of an operator algebra from its positive cone state structure).

Having settled the problem of uniqueness, the remaining task is to determine the action of emulats on wedge-localized
multi-particle vectors and to obtain explicit formulas for their particle formfactors.
These problems have been solved in case the domains of emulats are invariant under translations; in that case they
possess a~Fourier transform~\cite{BBS}.
This requirement is extremely restrictive and is only compatible with $d=1+1$ elastic two-particle scattering matrices of
integrable models\footnote{This statement, which I owe to Michael Karowski, is slightly stronger than that in~\cite{BBS}
in that that higher elastic amplitudes are combinatorial products of two-particle scattering functions, i.e.~the only
solutions are the factorizing models.}; in fact it should be considered as the \textit{foundational definition of
integrability of QFT in terms of properties of wedge-localized generator}~\cite{integrable}.

Since the action of emulats on particle states is quite complicated and still in a~conjectural stage, we will return to
this problem after explaining some more notation which is useful for formulating and proving the crossing identity in
connection with its KMS counterpart.
It will be helpful to the reader to recall how these properties have been derived in the integrable case.

For integrable models the wedge duality requirement leads to a~unique solution (the Zamo\-lod\-chi\-kov--Faddeev algebra),
whereas for the general non-integrable case we will present arguments which, together with some hindsight from the
integrable case, determine the action of emulates on particle states.
The main additional assumption is that the only way in which the interaction enters this construction of bijections is
through the $S$-matrix\footnote{A very reasonable assumption indeed because this is the only interaction-dependent object
which enters as a~relative modular invariant the modular theory for wedge localization.}.
With this assumption the form of the action of the operators $A_{\mathcal{A}(W)}$ on multiparticle states is
f\/ixed.
The ultimate check of its correctness through the verif\/ication of wedge duality is left to future investigations.

Whereas domains of emulats in the integrable case are translation invariant~\cite{BBS}, the only domain property which
is \textit{always} preserved in the general case is the invariance under the subgroup of those Poincar\'{e}
transformations which leave $W$ invariant.
In contrast to QM, for which integrability occurs in any dimension, integrability in QFT is restricted to $d=1+1$
factorizing models~\cite{integrable}.

A basic fact used in the derivation of the crossing identity, including its analytic properties which are necessary in
order to return to the physical boundary, is the \textit{cyclic KMS property.} For three operators connected with the
interacting algebra~$\mathcal{A}(W)$, one being from the algebra and two being emulates of incoming
operators\footnote{There exists also a~``free'' KMS identity in which~$B$ is replaced~by
$(B)_{\mathcal{A}_{\rm in}(W)}$ so everything refers to the algebra $\mathcal{A}_{\rm in}(W)$. The derivation of the
corresponding crossing identity is rather simple~\cite{integrable} and its use is limited to problems of writing
iterating f\/ields as a~series of Wick-ordered product of free f\/ields.}, it reads
\begin{gather}
  \big\langle 0|BA_{\mathcal{A}(W)}^{(1)}A_{\mathcal{A}(W)}^{(2)} |0\big\rangle
\overset{{\rm KMS}(\mathcal{A}(W))}{=}\big\langle 0|A_{\mathcal{A} (W)}^{(2)}\Delta BA_{\mathcal{A}(W)}^{(1)}|0\big\rangle
\label{M}
\\
  A^{(1)}\equiv:A(f_{1})\cdots A(f_{k}):, \qquad A_{\rm in}^{(2)}\equiv:A(f_{k+1})\cdots A(f_{n}):,
\qquad
\supp f_{i}\in W,
\nonumber
\end{gather}
where in the second line the operators were specialized to Wick-ordered products of smeared free f\/ields $A(f)$ which are
then emulated within~$\mathcal{A}(W)$. Their use is necessary in order to convert the KMS relation for \textit{fields}
af\/f\/iliated with $\mathcal{A}(W)$ into an identity of \textit{particle formfactors} of the operator $B\in\mathcal{A}(W)$.
If the bijective image acts on the vacuum, the subscript $\mathcal{A}(W)$ for the emulats can be
omitted and the resulting Wick-ordered product applied to the vacuum describe a~multi-particle state in $\hat{f}_{i}$
momentum space wave functions.
The roof on top of~$f$ denotes the wave function which results from the forward mass shell restriction of the Fourier
transform of $W$-supported test function.
The result are wave functions in a~Hilbert space of the graph norm $(\hat{f},(\Delta+1) \hat{f})$
which forces them to be analytic in the strip $0<\operatorname{Im}\theta<\pi$.

The derivation of the crossing relation requires to compute the formfactor of the emula\-te~$A_{\mathcal{A}(W)}^{(1)}$
between $W$-localized particle states and a~general $W$-localized state.
For simplicity of notation we specialize to $d=1+1$ in which case neither the wedge has a~transverse extension nor the
mass-shell momenta have a~transverse component; particles are then characterized by their rapidity.
Using the analytic properties of the wave functions which connect the complex conjugate of the antiparticle wave
function with the $i\pi$ boundary value of the particle wave function, one obtains
\begin{gather}
  \int\cdots \int\hat{f}_{1}(\theta_{1})\cdots \hat{f}_{1}(\theta_{n})F^{(k)} (\theta_{1},\dots,\theta_{n})d\theta_{1}\cdots
d\theta_{n}=0
\label{em1}
\end{gather}
with
\begin{gather*}
\begin{split}
&  F^{(k)}(\theta_{1},\dots ,\theta_{n})= \big\langle 0 \big\vert BA_{\mathcal{A}
(W)}^{(1)}(\theta_{1},\dots ,\theta_{k}) \big\vert \theta_{k+1},\dots ,\theta_{n} \big\rangle_{\rm in}\\
& \hphantom{F^{(k)}(\theta_{1},\dots ,\theta_{n})=}{}
- {}_{\rm out} \big\langle
\bar{\theta}_{k+1},\dots,\bar{\theta}_{n} \big\vert \Delta^{\frac{1}{2}}B \big\vert
\theta_{1},\dots,\theta_{k}\big\rangle_{\rm in}.
\end{split}
\end{gather*}
Here $\Delta^{\frac{1}{2}}$ of $\Delta$ was used to re-convert the antiparticle wave functions in the outgoing bra
vector back into the original particle wave functions.
The vanishing of $F^{(k)}$ is a~crossing relation which is certainly suf\/f\/icient for the validity of~\eqref{M}, but it
does not have the expected standard form which would result if we could omit the emulation subscript (in which case one
obtains the vacuum to $n$-particle matrix element of~$B$).
This is not allowed in the presence of interactions.
In the following we will show how integrable models solve this problem before we return to the general case.

In the integrable case \cite{BFKZ} the matrix-elements $\langle
0\vert B\vert \theta_{1},\dots,\theta_{n} \rangle $ are
meromorphic functions in the rapidities (not in the invariant Mandelstam
variables!). In that case there exists besides the degeneracy under statistics
exchange of $\theta s$ also the possibility of a \textit{nontrivial exchange
via analytic continuation}. In that case an analytic transposition of adjacent
$\theta s$ produces an $S(\theta_{i}-\theta_{i+1})$ factor, where~$S$ is the
scattering function of the model (the two-particle $S$-matrix from which all
higher elastic $S$-matrices are given in terms of a product formula)~\cite{BFKZ}. In order to distinguish between the analytic and the statistics
ordering change let us introduce the following notation
\begin{gather*}
 \langle 0 \vert B \vert \theta_{1},\dots,\theta_{n} \rangle
= \langle 0 \vert B \vert \theta_{1},\dots,\theta_{n} \rangle
_{\rm in}\equiv \langle 0 \vert B \vert \theta_{p_{1}},\dots,\theta
_{p_{n}} \rangle _{\rm in}, \qquad
\theta_{1}      >\theta_{2}>\cdots >\theta_{n},
\end{gather*}
where the formfactors with the subscript ${\rm in}$ obey the rules of statistics
degeneracy whereas the natural order on the left hand side requires analytic
continuations of the formfactor.

It has been observed in~\cite{AOP} and extended in~\cite{Lech} that the
analytic exchange can be encoded into the Zamolodchikov--Faddeev algebra
\begin{gather*}
   Z^{\ast}(\theta_{1})\cdots Z^{\ast}(\theta_{n}) \vert 0 \rangle
= \vert \theta_{1},\dots,\theta_{n} \rangle _{\rm in}, \qquad \theta_{1}
>\cdots >\theta_{n},\\
   Z(\theta)Z^{\ast}(\theta^{\prime})=\delta(\theta-\theta^{\prime}
)+S(\theta-\theta^{\prime}+i\pi)Z(\theta^{\prime})Z(\theta), \\
   Z^{\ast}(\theta)Z^{\ast}(\theta^{\prime})=S(\theta-\theta^{\prime})Z^{\ast
}(\theta^{\prime})Z^{\ast}(\theta),
\end{gather*}
where the $Z^{\prime}s$ are related to the emulate operators as
\begin{gather*}
A_{\rm in}(f)_{\mathcal{A}(W)}=\int_{\partial C}Z^{\ast}(\theta)e^{ip(\theta
)x}\hat{f}(\theta)d\theta, \qquad C=(0,i\pi) \ \ \text{strip}, \qquad Z(\theta)=Z^{\ast}(\theta+i\pi).
\end{gather*}
This leads in particular to
\begin{gather*}
 \langle 0 \vert B \vert \theta_{2},\dots,\theta_{k},\theta
_{1},\theta_{k+1},\dots,\theta_{n} \rangle     =S_{\rm gs} \langle
0 \vert B \vert \theta_{1},\dots,\theta_{n} \rangle _{\rm in},\\
S_{\rm gs}=
\prod\limits_{l=2}^{k}
S(\theta_{l}-\theta_{1}), \qquad 
\theta_{1}     >\cdots >\theta_{n}.
\end{gather*}
We will refer to $S_{\rm gs}$ as the ``grazing shot $S$-matrix'', the reason being
that it describes a scattering process in which scattering $\theta_{1}$ through the $\theta$-cluster $\theta_{2},\dots,\theta_{k}$ does not cause any
change within the cluster.

We now look for an analog of this construction in the generals case.
The main complication results from the presence of all inelastic
threshold singularities of multiparticle scattering which enter all analytic
order changes. So the f\/irst question is whether there exists an analog of the
above grazing shot $S$-matrix in the general case. For this purpose it
is helpful to rewrite the above integrable~$S_{\rm gs}$ into an
expression which only involves the full $S$-matrices. It is clear that in the
above example this leads
\begin{gather*}
S_{\rm gr}(\theta_{1}; \theta_{1},\dots,\theta_{k})=S(\theta_{2},\dots,\theta_{k})^{\ast
}S(\theta_{1},\dots,\theta_{k})
\end{gather*}
with the  $S$ being the full $S$-matrices of $k$ respectively $k-1$
particles does the job. In case the two-particle scattering matrix is not just
a~scattering function but rather a matrix of scattering functions, one has to
use the Yang--Baxter relation in order to cancel all interactions within the
$k-1$ cluster  $\theta_{2},\dots,\theta_{k}$.

This idea suggest to def\/ine a general grazing shot $S$-matrix as
\begin{gather}
S_{\rm gs}^{(m,n)}(\chi|\theta_{1};\theta)   \equiv\sum\limits_{l}\int\cdots \int d\vartheta_{1}\cdots d\vartheta_{m} \langle
\chi_{1}\cdots \chi_{m}|S^{\ast}|\vartheta_{1},\dots, \vartheta_{l} \rangle
\nonumber
\\
\hphantom{S_{\rm gs}^{(m,n)}(\chi|\theta_{1};\theta)   \equiv}{}
  \times \langle \theta_{1},\vartheta_{1},\dots,\vartheta_{l}|S|\theta_{1},\theta_{2},\dots,\theta_{k} \rangle.\label{gs}
\end{gather}
The~$\chi$ represents the $\chi=\chi_{1},\dots,\chi_{m}$ component of a~scattering process in which the grazing
shot ``bullet'' $\theta_{1}$ impinges on a~$k-1$ particle~$\theta$-cluster consisting of $\theta_{2},\dots,\theta_{k}$
particles.
Here the sum extends over all intermediate particles with energetically accessible thresholds, i.e.~the number of
intermediate open~$l$-channels increase with the initial energy.
The matrix elements of the creation part of an emulate sandwiched between two multi-particle states can directly be
written in terms of the grazing shot $S$-matrix as
\begin{gather*}
_{\rm in} \langle \chi_{1},\dots,\chi_{m} \vert Z^{\ast}(\theta)_{\mathcal{A} (W)} \vert
\theta_{1};\theta_{2},\dots,\theta_{n} \rangle_{\rm in} =S_{\rm gs}^{(m,n)}(\chi,\theta_{1};\theta),
\end{gather*}
where the $Z^{\ast}$ denotes the previously def\/ined creation part of the
emulat $A_{\rm in}(x)_{\mathcal{A}(W)}$.
Once the annihilation operator has been commuted through to its natural position, it annihilats the next particle on
the right and contributes a~delta contraction.
This procedure may be interpreted as a~generalization of Wick ordering to interacting emulats.

The indicated idea to generalize the algebraic structure of integrable models
by extending the concept of a grazing shot $S$-matrix is very speculative. But
without knowing more about the structure of emulats it is not possible to
generalize the $S$-matrix based ideas which f\/inally led to the mathematical
existence proof for integrable models of QFT.

Note that the previous arguments which led to the crossing relation%
\begin{gather}
 \langle 0 \vert B \vert \theta_{1},\dots,\theta_{k},\theta
_{k+1},\dots,\theta_{n} \rangle _{\rm in}= {}_{\rm out} \langle \bar{\theta}
_{k+1},\dots,\bar{\theta}_{n} \vert U(\Lambda_{W_{(0,1)}}(\pi i))B \vert
\theta_{1},\dots,\theta_{k} \rangle _{\rm in},\nonumber\\
 \langle 0 \vert B \vert p_{1},\dots, p_{k},p_{k+1}, \dots,p_{n}\rangle _{\rm in}={} _{\rm out}\langle -\bar{p}_{k+1},\dots,-\bar{p}
_{n} \vert U(\Lambda_{W_{(0,1)}}(\pi i))B \vert p_{1},\dots,p_{k} \rangle _{\rm in},\nonumber\\
  B\in\mathcal{A(O}), \qquad \mathcal{O}\subseteq W_{(0,1)}, \qquad \bar{\theta}= \text{antiparticle~of} \ \theta, \qquad \theta_{1}>\cdots >\theta_{n}\label{cro}
\end{gather}
are only valid in case of the natural order, in particular they do not depend
on the form of the emulat operators. The form of the crossing relation for
other orderings has to be computed in terms of analytic continuation which
leads to interaction-dependent modif\/ications whose form is only known for
integrable models. The second line in~(\ref{cro}) is the usual momentum space
form of the crossing relation; the $\theta$-ordering passes to the $p$-ordering.

The ordering dependence of the crossing relation receives additional
support from the Haag--Ruelle derivation of the LSZ reduction formalism~\cite{B-S}. There are indeed threshold modif\/ications from overlapping wave
functions which invalidate the derivation LSZ reduction formalism to such cases~\cite{Buch}.

The connections between the restriction of the LSZ scattering theory with the
idea of analytic $\theta$-changes looks very interesting and should be pursued
further.
They indicate the possible existence of deep unexplored connections between analytic threshold singularities and
algebraic emulats.
In this context the concept of emulation should be seen (as indicated in the title of this section) as a~generalization
of the functorial relation between the Wigner representation theoretical particle setting and the net structure of
interaction-free QFTs
\begin{gather*}
\text{functorial~relation}\overset{\rm interaction}{\longrightarrow} \text{emulation}.
\end{gather*}
In both cases the important role is played by modular localization theory.
Its use in the presence of (any) interactions is however a double-edged
sword. Without it QFT would loose its foundational character expressed in its
many structural theorems which have no counterpart in~QM. But it is precisely
this fundamental aspect which leads to the coupling of all states with the
same superselected charges which renders quantum mechanical approximation
methods rather inutil. Looked upon from the side of QM and its operator
methods (establishing selfadjointness, spectral resolutions, \dots) QFT appears
like a realization of ``Murphy's law'': \textit{everything which is not
forbidden to couple $($subject to the validity of the superselection rules$)$
actually does couple}. Only if one learns the appropriate operator algebraic
methods this curse becomes a blessing. It is precisely the modular
localization property in the presence of (any) interaction which is behind the
derivation of all structural properties of QFT.

The relative simplicity of integrable models results from the rather easy
algebraic structure of its wedge-localized generators which in turn
results from the simplicity of the (possibly matrix-valued) elastic
scattering functions. This makes it possible to describe the wedge-generators
in terms of deformed free f\/ields~\cite{Le}. In this case Murphy's law only
applies to of\/f-shell correlation functions or compact localized operators;
they continue to couple to all states to which the superselection principle
allows them to couple. For this reason the proof of the nontriviality of
compact localized double cone algebras from intersection of wedge algebras is
quite an important achievement; the proof is based on the use of ``modular
nuclearity''~\cite{Lech}.

In the integrable case it leads to a representation of the permutation group~\cite{Lech} and the possibility to construct wedge generators for given
scattering function by ``deformations'' of free f\/ields~\cite{Le}. In general the
analytic exchange is path-dependent (ref\/lecting the inf\/luence of the inelastic
threshold cuts) and the action of emulats on particle states becomes much more
complicated. This situation is vaguely reminiscent of a $d=1+2$ Wightman theory
with braid group statistics~\cite{B-M} for which the Bargman--Wightman--Hall
analyticity domain~\cite{St-Wi} is not ``schlicht'' but contains cuts in the
physical spacetime region. It is an interesting question whether the
path-dependence of analytic ordering changes can be encoded into a group
structure which resembles that of an inf\/inite braid group representation.

\section{Resum\'{e}
and concluding remarks}

The main point of the present work is to introduce a~new Hilbert space setting for higher spin interaction, including
those which hitherto have been treated in the Krein space setting of gauge theory.
The idea came out of modular localization, a~concept which the author already introduced in the 90s and which a~decade
ago became the starting point of a~new project for rigorous constructions of integrable models.
The third subject of this work is the critique of string theory from the viewpoint of causal localization.
Although these two subjects were already treated in previous publications by the author, there are new interesting
observations about their relation to modular localization.

Modular localization theory  helps to recognize and analyze past failures.
Looking back at the $S$-matrix based on-shell construction attempts of the 60s with present hindsight, one realizes that
there was not much of a~chance at that time for understanding the subtle role of the particle crossing property in such project.

The predominant trial and error correcting computational oriented conduct of research was amazingly successful in
connection with the post wwII renormalized perturbation theory; but its success begun to wane in the $S$-matrix based
on-shell construction project as formulated by Stanley Mandelstam; in particular the conceptual origin of the particle
crossing property remained outside its range.

In Section~\ref{Section2} it was shown that the recognition of some of conceptual errors in the dual model and ST leads to
prof\/itable new insights.
The most intriguing misunderstanding which led to the dual model and ST was referred to in Section~\ref{Section2} as the picture
puzzle situation.
It is based in the curious observation that there exists an irreducible operator algebra which carries a~positive energy
representation with a~discrete $(m,s)$ particle spectrum in $d=10$ spacetime dimensions; the famous superstring
representation.
With a~little more forensic work one notices that it is the only known solution of a~problem formulated 1932 by Majorana
(in analogy to the ${\rm O}(4,2)$ description of the hydrogen spectrum): construct an inf\/inite component purely discrete
positive energy representation of the Poincar\'{e} group (inf\/inite component f\/ield equation) from an irreducible
operator algebra.
Neither Majorana nor the
group of  physicists who during the 60s studied ``dynamical
groups'' tried to embed the Lorentz group into a larger noncompact group (the
``dynamical group'' project of Barut, Fronsdal, Kleinert, \dots,~\cite{To}) found a
solution.
The connection between the $d=10$ component supersymmetric chiral model with the positive energy supersting representation
of the Poincar\'{e} group provided the only known solution of this group theoretic problem.
Its misreading as a~solution of an $S$-matrix problem  of a stringlocal
object in spacetime is a result of a misunderstanding which in this paper was
referred to as a ``picture puzzle''.
Brower's theorem~\cite{Brower} is a~pure group theoretical kinematical conclusion, it has no bearing on the scattering
theory of particles.

In Section~\ref{Section2}
it was also shown that such misreading of mathematical facts is not limited to string theory but also af\/fects surrounding areas.
The AdS-CFT correspondence is certainly a~mathematical fact but, its \textit{physical} use by Maldacena is the result of
overlooking the causality issue~\cite{Mal}.
Relations between QFTs in dif\/ferent spacetime dimensions (with the exception of the holographic projection onto
null-surfaces) violate the causal completion property which is an indispensable part of causality (the timelike
counterpart of Einstein's spacelike causal independence).
One may use such isomorphic relations between local nets in dif\/ferent spacetimes for calculational purposes (certain
calculations on the unphysical side may be simpler) but the interpretation has to be done on the physical side.
In terms of the modular localization
property this refers to the possible mismatch between the inner
approximation $\mathcal{A}(\mathcal{O})=\cup _{\mathcal{D}\subset \mathcal{O}}\mathcal{A}(\mathcal{D})$ by
unions of small double cones~$D$ and the outer approximation in terms of
wedges $\mathcal{A}(\mathcal{O}^{\prime \prime})=\cap _{W\supset \mathcal{O}}\mathcal{A}(W)$.
This also af\/fects the alleged validity of the
quasiclassical Kaluza--Klein dimensional reduction in full QFT and other popular games with extra dimensions.

In most cases the incorrect conclusions result from the
belief that quantum degree of freedom issues can be dealt with in
quasiclassical approximations. Any attempt to prove such incorrect ideas of
QFT in terms of correlation functions (instead of manipulating
Lagrangians) would have failed.

\looseness=-1
The once very successful approach to particle physics, which consisted in moving ahead on a~pure computational track~by
trial and error without precise conceptual investments and guidance, seem to have lost its momentum with the discovery
of the Standard Model which us the f\/irst successful project to describe electroweak interactions together with the strong
nuclear forces within a~framework of nonabelian gauge theory.
In this setting the Higgs mechanism played the role of relating massive vector mesons to their massless counterparts.
Early criticism of this idea disappeared in the maelstrom of time and gave way to a~complete stagnation which is manifested in the fact
that despite theoretical shortcomings this mechanism remained unchanged for more than 4 decades.

Instead of pursuing the serious objections of the f\/irst years after the appearance of the ``Higgs mechanism'', Big
Science has used it for its justif\/ication by declaring the Higgs mechanism to be this most important
discovery of this century.
The situation is aggravated by the fact that the small community of theoreticians dedicated to foundational research
(which shares most of the critical view of this paper) has resigned and turned away from ongoing problems of particle
theory.
This led to a~deep schism which makes it even more dif\/f\/icult to get out of the present situation.

The SLF setting of $s\geq1$ renormalizable perturbation theory in Hilbert space does not only shed a~quite dif\/ferent
light on the issue of the ``Higgs mechanism'', but also suggests a~precise def\/inition of conf\/inement in terms of the
vanishing of all correlations in which stringlocal zero mass f\/ields (gluons) or stringlocal quarks appear (except
$q$-$\bar{q}$ pairs with a~f\/inite connecting string) so that apart from such pairs only pointlike generated observables
survive.
It also suggests a perturbative proof based on generatlzations of Yennie--Frautschi--Suura type
perturbative calculations.

As in earlier times, progress in particle theory is not possible without removing incorrect ideas of the past and seeing
problems in a~in a~new light.
What is however dif\/ferent is that in earlier times (the times of Pauli, Feynman, Landau, Lehmann, Jost, \dots) the
inf\/luence of ``Big Science'' on fundamental theoretical research was much smaller.
There was a~well developed ``Streikultur'' in which the formation of globalized  monocultures had no
place.

I do not have an answer to this problem, but I think that it is necessary to f\/ind one in order to preserve the important
role which particle physics played in the past.

\subsection*{Acknowledgements}

I am indebted to Bernard Kay for informing me about his attempts to solve the imbalance in the Maldacena conjecture~by
adding external degrees of freedom~\cite{Kay}.
My thanks go also to Raymond Stora for explaining some aspects of BRST perturbation theory of Yang--Mills f\/ields and for
taking much interest in the SLF Hilbert space setting.
I am indebted to Jens Mund for making his as yet unpublished work on problems of string-localization available.
Last not least I thank the referees of my paper; without their constructive help it would not have reached its present
form.

\pdfbookmark[1]{References}{ref}
\LastPageEnding


\begin{thebibliography}{99}
\footnotesize \itemsep=0pt

\bibitem{Aste} Aste A., Scharf~G., D\"utsch M.,
On gauge invariance and spontaneous symmetry breaking,
\href{http://dx.doi.org/10.1088/0305-4470/30/16/019}{\textit{J.~Phys.~A: Math. Gen.}} {\bf 30} (1997),
5785--5792,
\href{http://arxiv.org/abs/hep-th/9705216}{hep-th/9705216}.

\bibitem{BFKZ}
Babujian H., Fring A., Karowski M., Zapletal A., Exact form factors in
  integrable quantum f\/ield theories: the sine-{G}ordon model, \href{http://dx.doi.org/10.1016/S0550-3213(98)00737-8}{\textit{Nuclear
  Phys.~B}} \textbf{538} (1999), 535--586, \href{http://arxiv.org/abs/hep-th/9805185}{hep-th/9805185}.

\bibitem{BPZ}
Belavin A.A., Polyakov A.M., Zamolodchikov A.B., Inf\/inite conformal symmetry in
  two-dimensional quantum f\/ield theory, \href{http://dx.doi.org/10.1016/0550-3213(84)90052-X}{\textit{Nuclear Phys.~B}} \textbf{241}
  (1984), 333--380.

\bibitem{To}
Bogolyubov N.N., Logunov A., Oksak A.I., Todorov I.T., General principles of
  quantum f\/ield theory, \textit{Ma\-the\-matical Physics and Applied Mathematics},
  Vol.~10, Kluwer Academic Publishers, Dordrecht, 1990.

\bibitem{Borchers}
Borchers H.-J., Field operators as {$C^{\infty }$} functions in spacelike
  directions, \href{http://dx.doi.org/10.1007/BF02749678}{\textit{Nuovo Cimento}} \textbf{33} (1964), 1600--1613.

\bibitem{Bo}
Borchers H.-J., On revolutionizing quantum f\/ield theory with {T}omita's modular
  theory, \href{http://dx.doi.org/10.1063/1.533323}{\textit{J.~Math. Phys.}} \textbf{41} (2000), 3604--3673.

\bibitem{BBS}
Borchers H.-J., Buchholz D., Schroer B., Polarization-free generators and the
  {$S$}-matrix, \href{http://dx.doi.org/10.1007/s002200100411}{\textit{Comm. Math. Phys.}} \textbf{219} (2001), 125--140,
  \href{http://arxiv.org/abs/hep-th/0003243}{hep-th/0003243}.

\bibitem{BEG}
Bros J., Epstein H., Glaser V., A proof of the crossing property for
  two-particle amplitudes in general quantum f\/ield theory, \href{http://dx.doi.org/10.1007/BF01646307}{\textit{Comm. Math.
  Phys.}} \textbf{1} (1965), 240--264.

\bibitem{B-M}
Bros J., Mund J., Braid group statistics implies scattering in
  three-dimensional local quantum physics, \href{http://dx.doi.org/10.1007/s00220-012-1560-6}{\textit{Comm. Math. Phys.}}
  \textbf{315} (2012), 465--488, \href{http://arxiv.org/abs/1112.5785}{arXiv:1112.5785}.

\bibitem{Brower}
Brower R.C., Spectrum-generating algebra and no-ghost theorem for the dual
  model, \href{http://dx.doi.org/10.1103/PhysRevD.6.1655}{\textit{Phys. Rev.~D}} \textbf{6} (1972), 1655--1662.

\bibitem{BGL}
Brunetti R., Guido D., Longo R., Modular localization and {W}igner particles,
  \href{http://dx.doi.org/10.1142/S0129055X02001387}{\textit{Rev. Math. Phys.}} \textbf{14} (2002), 759--785,
  \href{http://arxiv.org/abs/math-ph/0203021}{math-ph/0203021}.


\bibitem{Buch}
Buchholz D., Haag--{R}uelle approximation of collision states, \href{http://dx.doi.org/10.1007/BF01645982}{\textit{Comm.
  Math. Phys.}} \textbf{36} (1974), 243--253.

\bibitem{Bu}
Buchholz D., New light on infrared problems: sectors, statistics, spectrum and
  all that, \href{http://arxiv.org/abs/1301.2516}{arXiv:1301.2516}.

\bibitem{ABL}
Buchholz D., D'Antoni C., Longo R., Nuclear maps and modular structures.
  {I}.~{G}eneral properties, \href{http://dx.doi.org/10.1016/0022-1236(90)90104-S}{\textit{J.~Funct. Anal.}} \textbf{88} (1990),
  233--250.

\bibitem{Bu-Fr}
Buchholz D., Fredenhagen K., Locality and the structure of particle states,
  \href{http://dx.doi.org/10.1007/BF01208370}{\textit{Comm. Math. Phys.}} \textbf{84} (1982), 1--54.

\bibitem{BMT}
Buchholz D., Mack G., Todorov I., The current algebra on the circle as a germ
  of local f\/ield theories, \href{http://dx.doi.org/10.1016/0920-5632(88)90367-2}{\textit{Nuclear Phys.~B Proc. Suppl.}} \textbf{5B}
  (1988), 20--56.

\bibitem{Bu-So}
Buchholz D., Solveen C., Unruh ef\/fect and the concept of temperature,
  \href{http://dx.doi.org/10.1088/0264-9381/30/8/085011}{\textit{Classical Quantum Gravity}} \textbf{30} (2013), 085011, 9~pages,
  \href{http://arxiv.org/abs/1212.2409}{arXiv:1212.2409}.

\bibitem{B-S}
Buchholz D., Summers S.J., Scattering in relativistic quantum f\/ield theory:
  fundamental concepts and tools, \href{http://arxiv.org/abs/math-ph/0509047}{math-ph/0509047}.

\bibitem{Bu-Wi}
Buchholz D., Wichmann E.H., Causal independence and the energy-level density of
  states in local quantum f\/ield theory, \href{http://dx.doi.org/10.1007/BF01454978}{\textit{Comm. Math. Phys.}} \textbf{106}
  (1986), 321--344.

\bibitem{Fub}
Di~Vecchia P., The birth of string theory, in String theory and fundamental
  interactions, \href{http://dx.doi.org/10.1007/978-3-540-74233-3_4}{\textit{Lecture Notes in Phys.}}, Vol.~737, Springer, Berlin,
  2008, 59--118, \href{http://arxiv.org/abs/0704.0101}{arXiv:0704.0101}.

\bibitem{Do-Xu}
Dong C., Xu F., Conformal nets associated with lattices and their orbifolds,
  \href{http://dx.doi.org/10.1016/j.aim.2005.08.009}{\textit{Adv. Math.}} \textbf{206} (2006), 279--306, \href{http://arxiv.org/abs/math.OA/0411499}{math.OA/0411499}.

\bibitem{Du-Ja}
Duncan A., Janssen M., Pascual Jordan's resolution of the conundrum of the
  wave-particle duality of light, \href{http://dx.doi.org/10.1016/j.shpsb.2008.04.005}{\textit{Stud. Hist. Philos. Sci.~B Stud.
  Hist. Philos. Modern Phys.}} \textbf{39} (2008), 634--666, \href{http://arxiv.org/abs/0709.3812}{arXiv:0709.3812}.

\bibitem{Garcia}
D\"utsch M., Gracia-Bond\'{\i}a J.M., Scheck F., V\'arilly J.C., Quantum gauge
  models without (classical) Higgs mechanism, \href{http://dx.doi.org/10.1140/epjc/s10052-010-1432-1}{\textit{Eur. Phys.~J.~C}}
  \textbf{69} (2010), 599--621, \href{http://arxiv.org/abs/1001.0932}{arXiv:1001.0932}.

\bibitem{E-G}
Epstein H., Glaser V., The role of locality in perturbation theory,
  \textit{Ann. Inst. H.~Poincar\'e~A} \textbf{19} (1973), 211--295.

\bibitem{S-E}
Ezawa H., Swieca J.A., Spontaneous breakdown of symmetries and zero-mass
  states, \href{http://dx.doi.org/10.1007/BF01646447}{\textit{Comm. Math. Phys.}} \textbf{5} (1967), 330--336.

\bibitem{Fa-Sc}
Fassarella L., Schroer B., Wigner particle theory and local quantum physics,
  \href{http://dx.doi.org/10.1088/0305-4470/35/43/311}{\textit{J.~Phys.~A: Math. Gen.}} \textbf{35} (2002), 9123--9164,
  \href{http://arxiv.org/abs/hep-th/0112168}{hep-th/0112168}.

\bibitem{Haag}
Haag R., Local quantum physics. Fields, particles, algebras, 2nd ed., \href{http://dx.doi.org/10.1007/978-3-642-61458-3}{\textit{Texts and
  Monographs in Physics}}, Springer-Verlag, Berlin, 1996.

\bibitem{H-S}
Haag R., Schroer B., Postulates of quantum f\/ield theory, \href{http://dx.doi.org/10.1063/1.1703797}{\textit{J.~Math.
  Phys.}} \textbf{3} (1962), 248--256.

\bibitem{H-Sw}
Haag R., Swieca J.A., When does a quantum f\/ield theory describe particles?,
  \href{http://dx.doi.org/10.1007/BF01645906}{\textit{Comm. Math. Phys.}} \textbf{1} (1965), 308--320.

\bibitem{Ho-Wa}
Hollands S., Wald R.M., Quantum f\/ield theory is not merely quantum mechanics
  applied to low energy ef\/fective degrees of freedom, \href{http://dx.doi.org/10.1023/B:GERG.0000048980.00020.9a}{\textit{Gen. Relativity
  Gravitation}} \textbf{36} (2004), 2595--2603, \href{http://arxiv.org/abs/gr-qc/0405082}{gr-qc/0405082}.

\bibitem{Jaffe}
Jaf\/fe A.M., High-energy behavior in quantum f\/ield theory. I.~Strictly
  localizable f\/ields, \href{http://dx.doi.org/10.1103/PhysRev.158.1454}{\textit{Phys. Rev.}} \textbf{158} (1967), 1454--1461.

\bibitem{Jost}
Jost R., TCP-Invarianz der Streumatrix und interpolierende Felder,
  \href{http://dx.doi.org/10.5169/seals-113357}{\textit{Helvetica Phys. Acta}} \textbf{36} (1963), 77--82.

\bibitem{K-W}
K{\"a}hler R., Wiesbrock H.-W., Modular theory and the reconstruction of
  four-dimensional quantum f\/ield theories, \href{http://dx.doi.org/10.1063/1.1327597}{\textit{J.~Math. Phys.}} \textbf{42}
  (2001), 74--86.

\bibitem{Ka-Lo}
Kawahigashi Y., Longo R., Local conformal nets arising from framed vertex
  operator algebras, \href{http://dx.doi.org/10.1016/j.aim.2005.11.003}{\textit{Adv. Math.}} \textbf{206} (2006), 729--751,
  \href{http://arxiv.org/abs/math.OA/0407263}{math.OA/0407263}.

\bibitem{Kay}
Kay B.S., Ort{\'{\i}}z L., Brick walls and {A}d{S}/{CFT}, \href{http://dx.doi.org/10.1007/s10714-014-1727-x}{\textit{Gen.
  Relativity Gravitation}} \textbf{46} (2014), 1727, 55~pages,
  \href{http://arxiv.org/abs/1111.6429}{arXiv:1111.6429}.

\bibitem{Lech}
Lechner G., An existence proof for interacting quantum f\/ield theories with a
  factorizing $S$-matrix, \href{http://dx.doi.org/10.1007/s00220-007-0381-5}{\textit{Comm. Math. Phys.}} \textbf{227} (2008),
  821--860, \href{http://arxiv.org/abs/math-ph/0601022}{math-ph/0601022}.

\bibitem{Le}
Lechner G., Deformations of quantum f\/ield theories and integrable models,
  \href{http://dx.doi.org/10.1007/s00220-011-1390-y}{\textit{Comm. Math. Phys.}} \textbf{312} (2012), 265--302, \href{http://arxiv.org/abs/1104.1948}{arXiv:1104.1948}.

\bibitem{Lowe}
Lowe D.A., Causal properties of free string f\/ield theory, \href{http://dx.doi.org/10.1016/0370-2693(94)91314-5}{\textit{Phys.
  Lett.~B}} \textbf{326} (1994), 223--230, \href{http://arxiv.org/abs/hep-th/9312107}{hep-th/9312107}.

\bibitem{Lo-Sw}
Lowenstein J.H., Swieca J.A., Quantum electrodynamics in two dimensions,
  \href{http://dx.doi.org/10.1016/0003-4916(71)90246-6}{\textit{Ann. Physics}} \textbf{68} (1971), 172--195.

\bibitem{Ma-Lu}
L{\"u}scher M., Mack G., Global conformal invariance in quantum f\/ield theory,
  \href{http://dx.doi.org/10.1007/BF01608988}{\textit{Comm. Math. Phys.}} \textbf{41} (1975), 203--234.

\bibitem{Mack2}
Mack G., $D$-independent representation of conformal f\/ield theories in $D$
  dimensions via transformation to auxiliary dual resonance models. Scalar
  amplitudes, \href{http://arxiv.org/abs/0907.2407}{arXiv:0907.2407}.

\bibitem{Mack1}
Mack G., {$D$}-dimensional conformal f\/ield theories with anomalous dimensions
  as dual resonance models, 2009, \href{http://arxiv.org/abs/0909.1024}{arXiv:0909.1024}.

\bibitem{Maj}
Majorana E., Teoria relativistica di particelle con momentum internisico
  arbitrario, \href{http://dx.doi.org/10.1007/BF02959557}{\textit{Nuovo Cimento}} \textbf{9} (1932), 335--344.

\bibitem{Mal}
Maldacena J., The illusion of gravity, \href{http://dx.doi.org/10.1038/scientificamerican1105-56}{\textit{Sci. Amer.}} \textbf{293} (2005),
  no.~5, 56--63.

\bibitem{Mandel}
Mandelstam S., Determination of the pion-nucleon scattering amplitude from
  dispersion relations and unitarity. {G}eneral theory, \href{http://dx.doi.org/10.1103/PhysRev.112.1344}{\textit{Phys. Rev.}}
  \textbf{112} (1958), 1344--1360.

\bibitem{Mandel+}
Mandelstam S., Analytic properties of transition amplitudes in perturbation
  theory, \href{http://dx.doi.org/10.1103/PhysRev.115.1741}{\textit{Phys. Rev.}} \textbf{115} (1959), 1741--1751.

\bibitem{Man}
Mandelstam S., Feynman rules for electromagnetic and Yang--Mills f\/ields from the
  gauge-independent f\/ield-theoretic formalism, \href{http://dx.doi.org/10.1103/PhysRev.175.1580}{\textit{Phys. Rev.}} \textbf{175}
  (1968), 1580--1603.

\bibitem{Martinec}
Martinec E., The light cone in string theory, \href{http://dx.doi.org/10.1088/0264-9381/10/10/002}{\textit{Classical Quantum
  Gravity}} \textbf{10} (1993), L187--L192, \href{http://arxiv.org/abs/hep-th/9304037}{hep-th/9304037}.


\bibitem{Mund2}
Mund J., An algebraic {J}ost--{S}chroer theorem for massive theories,
  \href{http://dx.doi.org/10.1007/s00220-012-1546-4}{\textit{Comm. Math. Phys.}} \textbf{315} (2012), 445--464, \href{http://arxiv.org/abs/1012.1454}{arXiv:1012.1454}.


\bibitem{Jens}
Mund J., String-localized massive vector bosons without ghosts and indef\/inite
  metric: the example of massive QED, in preparation.

\bibitem{M-S}
Mund J., Schroer B., Renormalization theory of string-localized self-coupled
  massive vectormesons in Hilbert space, {i}n preparation.

\bibitem{MSY}
Mund J., Schroer B., Yngvason J., String-localized quantum f\/ields and modular
  localization, \href{http://dx.doi.org/10.1007/s00220-006-0067-4}{\textit{Comm. Math. Phys.}} \textbf{268} (2006), 621--672,
  \href{http://arxiv.org/abs/math-ph/0511042}{math-ph/0511042}.

\bibitem{Pla}
Plaschke M., Yngvason J., Massless, string localized quantum f\/ields for any
  helicity, \href{http://dx.doi.org/10.1063/1.3700765}{\textit{J.~Math. Phys.}} \textbf{53} (2012), 042301, 15~pages,
  \href{http://arxiv.org/abs/1111.5164}{arXiv:1111.5164}.

\bibitem{Polch}
Polchinski J., String theory.~I.~An introduction to the bosonic string,
  \textit{Cambridge Monographs on Mathema\-ti\-cal Physics}, Cambridge University Press,
  Cambridge, 1998.

\bibitem{Reh}
Rehren K.-H., Local quantum observables in the anti-de-{S}itter~-- conformal
  {QFT} correspondence, \href{http://dx.doi.org/10.1016/S0370-2693(00)01168-0}{\textit{Phys. Lett.~B}} \textbf{493} (2000), 383--388,
  \href{http://arxiv.org/abs/hep-th/0003120}{hep-th/0003120}.

\bibitem{Scharf}
Scharf G., Quantum gauge theories. A true ghost story, John Wiley \& Sons, New
  York, 2001.





\bibitem{S1}
Schroer B., Modular localization and the bootstrap-formfactor program,
  \href{http://dx.doi.org/10.1016/S0550-3213(97)00359-3}{\textit{Nuclear Phys.~B}} \textbf{499} (1997), 547--568,
  \href{http://arxiv.org/abs/hep-th/9702145}{hep-th/9702145}.

\bibitem{AOP}
Schroer B., Modular wedge localization and the {$d=1+1$} formfactor program,
  \href{http://dx.doi.org/10.1006/aphy.1999.5921}{\textit{Ann. Physics}} \textbf{275} (1999), 190--223, \href{http://arxiv.org/abs/hep-th/9712124}{hep-th/9712124}.

\bibitem{Swieca}
Schroer B., Jorge A.~Swieca's contributions to quantum f\/ield theory in the 60s
  and 70s and their relevance in present research, \href{http://dx.doi.org/10.1140/epjh/e2010-00004-1}{\textit{Eur. Phys. J.~H}}
  \textbf{35} (2010), 53--88, \href{http://arxiv.org/abs/0712.0371}{arXiv:0712.0371}.

\bibitem{interface}
Schroer B., Localization and the interface between quantum mechanics, quantum
  f\/ield theory and quantum gravity~{I}: the two antagonistic localizations and
  their asymptotic compatibility, \href{http://dx.doi.org/10.1016/j.shpsb.2010.03.003}{\textit{Stud. Hist. Philos. Sci.~B Stud.
  Hist. Philos. Modern Phys.}} \textbf{41} (2010), 104--127, \href{http://arxiv.org/abs/0912.2874}{arXiv:0912.2874}.

\bibitem{nonlocal}
Schroer B., Unexplored regions in QFT and the conceptual foundations of the
  Standard Model, \href{http://arxiv.org/abs/1006.3543}{arXiv:1006.3543}.

\bibitem{charge}
Schroer B., An alternative to the gauge theoretic setting, \href{http://dx.doi.org/10.1007/s10701-011-9567-y}{\textit{Found.
  Phys.}} \textbf{41} (2011), 1543--1568, \href{http://arxiv.org/abs/1012.0013}{arXiv:1012.0013}.

\bibitem{integrable}
Schroer B., Modular localization and the foundational origin of integrability,
\href{http://dx.doi.org/10.1007/s10701-013-9699-3}{\textit{Found.
  Phys.}} \textbf{43} (2013), 329--372,
  \href{http://arxiv.org/abs/1109.1212}{arXiv:1109.1212}.

\bibitem{causal}
Schroer B., Causality and dispersion relations and the role of the {$S$}-matrix
  in the ongoing research, \href{http://dx.doi.org/10.1007/s10701-012-9676-2}{\textit{Found. Phys.}} \textbf{42} (2012),
  1481--1522, \href{http://arxiv.org/abs/1107.1374}{arXiv:1107.1374}.

\bibitem{E-J}
Schroer B., The Einstein--Jordan conundrum and its relation to ongoing
  foundational research in local quantum physics, \href{http://dx.doi.org/10.1140/epjh/e2012-30059-x}{\textit{Eur. Phys.~J.~H}}
  \textbf{38} (2013), 137--173, \href{http://arxiv.org/abs/1101.0569}{arXiv:1101.0569}.



\bibitem{dark}
Schroer B., Dark matter and Wigner's third positive-energy representation
  class, \href{http://arxiv.org/abs/1306.3876}{arXiv:1306.3876}.


\bibitem{Schroer}
Schroer B., A Hilbert space setting for interacting higher spin f\/ields and the
  Higgs issue, \href{http://arxiv.org/abs/1407.0365}{arXiv:1407.0365}.

\bibitem{Sc-Sw}
Schroer B., Swieca J.A., Conformal transformations for quantized f\/ields,
  \href{http://dx.doi.org/10.1103/PhysRevD.10.480}{\textit{Phys. Rev.~D}} \textbf{10} (1974), 480--485.

\bibitem{schwinger}
Schwinger J., Trieste lectures, IAEA, Vienna, 1963.

\bibitem{Sewell}
Sewell G., Quantum f\/ields on manifolds: {PCT} and gravitationally induced
  thermal states, \href{http://dx.doi.org/10.1016/0003-4916(82)90285-8}{\textit{Ann. Physics}} \textbf{141} (1982), 201--224.

\bibitem{Sta}
Staskiewicz C.P., Die lokale Struktur abelscher Stromalgebren auf dem Kreis,
  Ph.D. Thesis, Freie Universit\"at Berlin, 1995.

\bibitem{St-Wi}
Streater R.F., Wightman A.S., P{CT}, spin and statistics, and all that, W. A.
  Benjamin, Inc., New York~-- Amsterdam, 1964.

\bibitem{Sw}
Swieca J.A., Charge screening and mass spectrum, \href{http://dx.doi.org/10.1103/PhysRevD.13.312}{\textit{Phys. Rev.~D}}
  \textbf{13} (1976), 312--314.

\bibitem{Unruh}
Unruh W.G., Notes on black-hole evaporation, \href{http://dx.doi.org/10.1103/PhysRevD.14.870}{\textit{Phys. Rev.~D}} \textbf{14}
  (1976), 870--892.

\bibitem{Ven}
Veneziano G., Construction of a crossing-simmetric, Regge-behaved amplitude for
  linearly rising trajectories, \href{http://dx.doi.org/10.1007/BF02824451}{\textit{Nuovo Cimento~A}} \textbf{57} (1968),
  190--197.

\bibitem{Weinbook}
Weinberg S., The quantum theory of f\/ields. I.~Foundations, Cambridge University
  Press, Cambridge, 2005.

\bibitem{YFS}
Yennie D., Frautschi S., Suura H., The infrared divergence phenomena and
  high-energy processes, \href{http://dx.doi.org/10.1016/0003-4916(61)90151-8}{\textit{Ann. Physics}} \textbf{13} (1961), 379--452.

\bibitem{Yold}
Yngvason J., Zero-mass inf\/inite spin representations of the {P}oincar\'e group
  and quantum f\/ield theory, \href{http://dx.doi.org/10.1007/BF01649432}{\textit{Comm. Math. Phys.}} \textbf{18} (1970),
  195--203.

\end{thebibliography}
\end{document}